\documentclass[11pt,a4paper]{JHEP3}
\usepackage{amssymb}
\usepackage{euscript}
\usepackage{epsfig,bbm,amssymb,euscript}

\def\b{{\beta}}
\def\d{\partial}
\def\1{{\bf 1}}
\def\x{{\bf x}}

\def\d{\partial}

\def\ln{{\rm ln}}
\def\a{\alpha}

\def\0{\nonumber}
\newcommand{\tU}{{\tilde U}}
\newcommand{\tX}{{\tilde X}}
\newcommand{\tN}{{\tilde N}}

\newcommand{\tV}{{\tilde V}}
\newcommand{\tE}{{\tilde E}}
\newcommand{\tT}{{\tilde T}}
\newcommand{\tS}{{\tilde S}}

\newcommand{\U}{{\mathcal U}}
\newcommand{\A}{{\mathcal A}}
\newcommand{\Q}{{\mathcal Q}}
\newcommand{\N}{{\mathcal N}}

\newcommand{\T}{{\mathcal T}}

\newcommand{\cS}{{\mathcal S}}

\newcommand{\Y}{{\mathcal Y}}
\newcommand{\V}{{\mathcal V}}
\newcommand\ES{\EuScript{S}}

\newcommand\ET{\EuScript{T}}

\newcommand\EX{\EuScript{X}}
\newcommand\EV{\EuScript{V}}

\newcommand\ER{\EuScript{R}}
\newcommand\EP{\EuScript{P}}
\newcommand\I{\mathbb{I}}
\def\beq{\begin{equation}}
\def\eeq{\end{equation}}
\def\eea{\end{eqnarray}}     
\def\bea{\begin{eqnarray}}

\def\e{{\rm e}}

\font\doppio=msbm10 at11pt
\newcommand{\RR}{\hbox {\doppio R}}

\preprint{SISSA/24/03/EP  \\ HU-EP-03/14 \\\tt hep-th/0304270}

\title{  Topics in String Field Theory}

\author { L. Bonora, C. Maccaferri, D. Mamone \\
International School for Advanced Studies (SISSA/ISAS)\\
Via Beirut 2--4, 34014 Trieste, Italy, and INFN, Sezione di
Trieste\\
E-mail:   \email{bonora@sissa.it}, \email{maccafer@sissa.it}, \email{mamone@sissa.it}}

\author{ M. Salizzoni \\
Humboldt Universit\"at zu Berlin \\
Institut f\"ur Physik \\
Newtonstrasse 15, Berlin, Germany \\
E-mail: \email{sali@physik.hu-berlin.de} }

\abstract{This review of bosonic string field theory is concentrated
on two main subjects. In the first part we revisit the construction
of the three string vertex and rederive the relevant Neumann
coefficients both for the matter and the ghost part following a
conformal field theory approach.
We use this formulation to solve the VSFT equation of motion
for the ghost sector. This part of the paper is based on a new method
which allows us to derive known results in a simpler way.
In the second part we concentrate on the solution of the VSFT
equation of motion for the matter part. We describe the construction
of the three strings vertex in the presence of a background $B$ field.
We determine a large family of lump solutions, illustrate their
interpretation as D--branes and study the low energy limit.
We show that in this limit the lump solutions flow toward the so--called
GMS solitons. }

\begin{document}

\section{Introduction}

Recently, as a consequence of the increasing interest in tachyon
condensation, String Field Theory (SFT) has received a renewed attention.
There is no doubt that the most complete description of tachyon
condensation and related phenomena has been given so far in the framework
of Witten's Open String Field Theory, \cite{W1}. This is not surprising,
since the study of tachyon condensation involves off--shell
calculations, and SFT is the natural framework where off--shell
analysis can be carried out.

All these developments can be described along the blueprint represented
by A.Sen's conjectures, \cite{Sen}. The latter can be summarized as
follows. Bosonic open  string theory in D=26 dimensions is quantized
on an unstable vacuum, an instability which manifests itself through the
existence of the open string tachyon. The effective tachyonic potential
has, beside the local maximum where the theory is quantized, a local
minimum. Sen's conjectures concern the nature of the theory around
this local minimum. First of all, the energy density difference between
the maximum and the minimum should exactly compensate for the
D25--brane tension characterizing the unstable vacuum: this is a
condition for the stability of the theory at the minimum.
Therefore the theory around the minimum should not contain any quantum
fluctuation pertaining to the original (unstable) theory. The minimum
should therefore correspond to an entirely new theory, the bosonic
closed string theory. If so, in the new theory one should be able to
explicitly find in particular all the classical solutions
characteristic of closed string theory, specifically the D25--brane
as well as all the lower dimensional D--branes.

The evidence that has been found for the above conjectures does not have a
uniform degree of accuracy and reliability, but it is enough to conclude
that they provide a correct description of tachyon condensation in
SFT. Especially elegant is the proof of the existence of solitonic
solutions in Vacuum String Field Theory (VSFT), the SFT version which is
believed to represent the theory near the minimum.

The aim of this review is not of giving a full account of the entire
subject of SFT and tachyon condensation. In this regards there are
already several reviews, \cite{ohmori,aref,RSZ1,WT}. We will rather
concentrate on some specific topics which are not covered in other
reviews. The first part of this article is devoted to the operator
formulation of SFT. The reason for this is that the latest
developments in VSFT and especially in supersymmetric SFT (see in particular
\cite{ohmori,aref}), have brought up aspects of the theory
that had not been analyzed in sufficient detail in the existing literature.
We refer in particular to the ghost structure of SFT and the relation
between the operator formulation and the (twisted) conformal field theory
formulation. To clarify this issue we extensively use the CFT interpretation
of SFT, advocated especially by \cite{leclair1,leclair2}.

The second part of this review is a synopsis
of D--branes in VSFT and noncommutative solitons. Our main purpose
is finding families of tachyonic lumps that can
consistently be interpreted as D--branes and studying their low energy
limit. We do so by introducing a constant background $B$ field, with
the purpose of smoothing out some singularities that appear in the low
energy limit when the $B$ field is absent. The result is rewarding:
we find a series of noncommutative solitons (the GMS solitons) that were
found some time ago by studying noncommutative effective field theories
of the tachyon.

\section{A summary of String Field Theory}

The open string field theory action proposed by E.Witten, \cite{W1}, years
ago is
\beq
{\cal S}(\Psi)= - \frac 1{g_0^2} \int\left(\frac 12 \Psi *Q\Psi +
\frac 13 \Psi *\Psi *\Psi\right)\label{sftaction}
\eeq
In this expression $\Psi$ is the string field, which can be understood
either as a classical functional of the open string configurations or as
a vector in the Fock space of states of the open string. We will consider
in the following the second point of view. In the field theory limit
it makes sense to represent it as a superposition of Fock space states
with ghost number 1, with coefficient represented by local fields,
\beq
|{\Psi}\rangle = (\phi(x)+ A_\mu (x) a_1^{\mu\dagger}+
\ldots) c_1|{0}\rangle\label{stringfield}
\eeq
The BRST charge $Q$ has the same form as in the first quantized string
theory. The star product of two string fields $\Psi_1, \Psi_2$ represents
the process of identifying the right half of the first string
with the left half of the second string and integrating over the
overlapping degrees of freedom, to produce a third string which
corresponds to $\Psi_1 * \Psi_2$. This can be done in various ways, either
using the classical string functionals (as in the original formulation), or
using the three string vertex (see below), or the conformal field theory
language \cite{leclair1}.
Finally the integration in (\ref{sftaction}) corresponds to bending the left
half of the string over the right half and integrating over the
corresponding degrees of freedom in such a way as to produce a number.
The following rules are obeyed
\bea
&&Q^2 =0 \0\\
&& \int Q \Psi = 0\0\\
&& (\Psi_1 *\Psi_2) *\Psi_3 = \Psi_1 * (\Psi_2 *\Psi_3)\0\\
&& Q(\Psi_1 * \Psi_2) = (Q \Psi_1)*\Psi_2 +
(-1)^{|\Psi_1|} \Psi_1 * (Q\Psi_2)\label{rules}
\eea
where $|\Psi|$ is the Grassmannality of the string field $\Psi$,
whic, for bosonic strings, coincides with the ghost number.
The action (\ref{sftaction}) is invariant under the BRST transformation
\beq
\delta \Psi = Q \Lambda + \Psi *\Lambda - \Lambda * \Psi\label{gaugetr}
\eeq
Finally, the ghost numbers of the various objects $Q, \Psi, \Lambda,
*, \int$ are $1,1,0,0,-3$, respectively.

Following these rules it is possible to explicitly compute the action
(\ref{sftaction}). For instance, in the low energy limit, where the
string field may be assumed to take the form (\ref{stringfield}),
the action becomes an integrated
function $F$ of an infinite series of local polynomials
(kinetic and potential terms)
of the fields involved in (\ref{stringfield}):
\beq
{\cal S}(\Psi) = \int d^{26}x F(\varphi_i,\d\varphi_i,...)
\label{effaction}
\eeq

\subsection{Vacuum string field theory}

The action (\ref{sftaction}) represents open string theory about the
trivial unstable vacuum $|\Psi_0\rangle=c_1|0\rangle$.
Vacuum string field theory (VSFT) is instead a version of Witten's open SFT
which is conjectured to correspond to the minimum of the tachyon potential.
As explained in the introduction at the minimum of the tachyon
potential a dramatic change occurs in the theory, which, corresponding
to the new vacuum, is expected to represent closed string
theory rather that the open string theory we started with. In particular,
this theory should host tachyonic lumps representing unstable D--branes
of any dimension less than 25, beside the original D25--brane.
Unfortunately we have been so
far unable to find an exact classical solution, say $|\Phi_0\rangle$,
representing the new vacuum. One can nevertheless guess the form taken
by the theory at the new minimum, see \cite{RSZ1}. The VSFT action
has the same form as (\ref{sftaction}), where the new string field is
still denoted by $\Psi$, the $*$ product is the same as in the previous
theory, while the BRST operator
$Q$ is replaced by a new one, usually denoted ${\cal Q}$, which is
characterized by universality and vanishing cohomology.
Relying on such general arguments, one can even deduce
a precise form of $\cal Q$ (\cite{GRSZ1},\cite{HK}, see also
\cite{HKw,Oku1,Oku2,RSZ4,RSZ5,Kishi}
and \cite{RSZ3,GT,KO,Moeller,David,Mukho,Sch1}),
\beq
{\cal {Q}} =  c_0 + \sum_{n>0} \,(-1)^n \,(c_{2n}+ c_{-2n})\label{calQ}
\eeq
Now, the equation of motion of VSFT is
\beq
{\cal Q} \Psi = - \Psi * \Psi\label{EOM}
\eeq
and nonperturbative solutions are looked for in the factorized form
\beq
\Psi= \Psi_m \otimes \Psi_g\label{ans}
\eeq
where $\Psi_g$ and $\Psi_m$ depend purely on ghost and matter
degrees of freedom, respectively. Then eq.(\ref{EOM}) splits into
\bea
 {\cal Q} \Psi_g & = & - \Psi_g * \Psi_g\label{EOMg}\\
\Psi_m & = & \Psi_m * \Psi_m\label{EOMm}
\eea

We will see later on how to compute solutions to both equations.
A solution to eq.(\ref{EOMg}) was calculated in \cite{GRSZ1,HKw}.
Various solutions of the matter part have been found in the
literature, \cite{RSZ1,HK,RSZ3,GT,KP,FKM,GRSZ2}.

\subsection{Organization of the paper. First part}

In the first part of this review we rederive the three strings vertex
coefficients by relying on the definition and the methods introduced
by \cite{leclair1,leclair2,Samu,Sue}. We do it first for the matter part
(section 3). In section 4 we derive the ghost
Neumann coefficients and in section 5 we concentrate on the
equation of motion of VSFT and look for matter--ghost factorized
solutions. We show how to rederive the solution for the ghost
part with a new method. Finally section 6 is meant as an introduction
to the second part of the paper.

\section{Three strings vertex and matter Neumann coefficients}

The three strings vertex \cite{W1,GJ1,GJ2} of Open String Field
Theory is given by
\beq
|V_3\rangle= \int d^{26}p_{(1)}d^{26}p_{(2)}d^{26}p_{(3)}
\delta^{26}(p_{(1)}+p_{(2)}+p_{(3)})\,{\rm exp}(-E)\,
|0,p\rangle_{123}\label{V3}
\eeq
where
\beq
E= \sum_{a,b=1}^3\left(\frac 12 \sum_{m,n\geq 1}\eta_{\mu\nu}
a_m^{(a)\mu\dagger}V_{mn}^{ab}
a_n^{(b)\nu\dagger} + \sum_{n\geq 1}\eta_{\mu\nu}p_{(a)}^{\mu}
V_{0n}^{ab}
a_n^{(b)\nu\dagger} +\frac 12 \eta_{\mu\nu}p_{(a)}^{\mu}V_{00}^{ab}
p_{(b)}^\nu\right) \label{E}
\eeq
Summation over the Lorentz indices $\mu,\nu=0,\ldots,25$
is understood and $\eta$ denotes the flat Lorentz metric.
The operators $ a_m^{(a)\mu},a_m^{(a)\mu\dagger}$ denote the non--zero
modes matter oscillators of the $a$--th string, which satisfy
\beq
[a_m^{(a)\mu},a_n^{(b)\nu\dagger}]=
\eta^{\mu\nu}\delta_{mn}\delta^{ab},
\quad\quad m,n\geq 1 \label{CCR}
\eeq
$p_{(r)}$ is the momentum of the $a$--th string and
$|0,p\rangle_{123}\equiv |p_{(1)}\rangle\otimes
|p_{(2)}\rangle\otimes |p_{(3)}\rangle$ is
the tensor product of the Fock vacuum
states relative to the three strings. $|p_{(a)}\rangle$ is
annihilated by
the annihilation
operators $a_m^{(a)\mu}$ and it is eigenstate of the momentum operator
$\hat p_{(a)}^\mu$
with eigenvalue $p_{(a)}^\mu$. The normalization is
\beq
\langle p_{(a)}|\, p'_{(b)}\rangle = \delta_{ab}\delta^{26}(p+p')\0
\eeq
The symbols $V_{nm}^{ab},V_{0m}^{ab},V_{00}^{ab}$ will denote
the coefficients computed in \cite{GJ1,GJ2}. We will use them
in the notation of Appendix A and B of \cite{RSZ2} and refer to them
as the {\it standard} ones. The notation $V_{MN}^{rs}$ for them
will also be used at times (with $M(N)$ denoting the couple
$\{0,m\}$ ($\{0,n\}$)) .

An important ingredient in the following are the $bpz$ transformation properties of
the oscillators
\beq
bpz(a_n^{(a)\mu}) = (-1)^{n+1} a_{-n}^{(a)\mu}\0
\eeq

Our purpose here is to discuss the definition and the properties of the
three strings vertex by exploiting as far as possible the definition
given in \cite{leclair1} for the Neumann coefficients. Remembering the
description of the
star product given in the previous section, the latter is obtained in the
following way. Let us consider three unit semidisks in the upper half
$z_a$ ($a=1,2,3$)
plane. Each one represents the string freely propagating in semicircles from the
origin (world-sheet time $\tau = -\infty$) to the unit circle $|z_a|=1$
($\tau =0$), where the interaction is supposed to take place. We map
each unit semidisk to a $120^\circ$ wedge of the complex plane
via the following conformal maps:
\beq
f_a(z_a)=\alpha^{2-a} f(z_a) \, ,\, a=1,2,3
\eeq
where
\beq
f(z)=\Big{(} \frac{1+iz}{1-iz}\Big{)} ^{\frac{2}{3}}
\eeq
Here $\alpha=e^{\frac{2\pi i}{3}}$ is one of the three third roots of
unity.
In this way the three semidisks are mapped to nonoverlapping (except
at the interaction points $z_a=i$) regions in such a way as to fill up a
unit disk centered at the origin. The curvature is zero everywhere
except at the center of the disk, which represents the common midpoint of
the three strings in interaction.

\begin{figure}[htbp]
    \hspace{-0.5cm}
\begin{center}
    \includegraphics[scale=0.5]{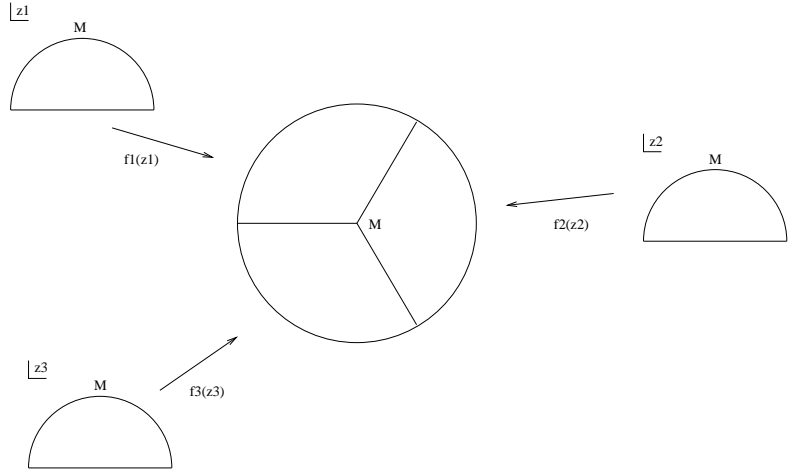}
    \end{center}
\caption{\emph{\small The conformal maps from the three unit semidisks
to the three-wedges unit disk}}
    \label{fig:A}
\end{figure}

The interaction vertex is defined by a correlation function on the disk in
the following way
\beq\label{cover}
\int\psi*\phi*\chi=\langle f_1\circ\psi(0)\, f_2\circ\phi(0)\, f_3\circ\chi(0)\rangle=\langle V_{123}|\psi\rangle_{1}|\phi\rangle_2|\chi\rangle_3
\eeq
Now we consider the string propagator
at two generic points of this disk. The Neumann coefficients $N^{ab}_{NM}$
are nothing but the Fourier modes of the propagator with respect to the
original coordinates $z_a$. We shall see that such Neumann coefficients
are related in a simple way to the standard three strings vertex coefficients.

Due to the qualitative difference between the $\a_{n>0}$ oscillators
and the zero modes $p$, the Neumann coefficients involving the latter
will be treated separately.

\subsection{Non--zero modes}

The Neumann coefficients $N_{mn}^{ab}$ are given by \cite{leclair1}
\beq
N_{mn}^{ab}=\langle V_{123}|\a_{-n}^{(a)}\a_{-m}^{(b)}|0\rangle_{123}=
-\frac{1}{nm}\oint\frac{dz}{2\pi i}\oint\frac{dw}{2\pi i}
\frac{1}{z^{n}}\frac{1}{w^{m}}f'_a(z)
\frac{1}{(f_a(z)-f_b(w))^2}f'_b(w)\label{neumann}
\eeq
where the contour integrals are understood around the origin.
It is easy to check that
\bea
N_{mn}^{ab}&=&N_{nm}^{ba}\nonumber\\
N_{mn}^{ab}&=&(-1)^{n+m}N_{mn}^{ba}\label{cycl}\\
N_{mn}^{ab}&=&N_{mn}^{a+1,b+1}\0
\eea
In the last equation the upper indices are defined mod 3.
\par
Let us consider the decomposition
\beq
N_{mn}^{ab}=\frac{1}{3\sqrt{nm}}{\Big{(}}E_{nm}+\bar{\a}^{a-b}U_{nm}+
\a^{a-b}\bar{U}_{nm}{\Big{)}}\label{decomp}
\eeq
After some algebra one gets
\bea
E_{nm}\!&=&\!\frac{-1}{\sqrt{nm}}\oint\frac{dz}{2\pi i}
\oint\frac{dw}{2\pi i}
\frac{1}{z^{n}}\frac{1}{w^{m}}
\Big{(}\frac{1}{(1+zw)^2}+\frac{1}{(z-w)^2}\Big{)}\\
\label{Anm}
U_{nm}\!&=&\!\frac{-1}{3\sqrt{nm}}\oint\frac{dz}{2\pi i}\oint\frac{dw}{2\pi i}
\frac{1}{z^{n}}\frac{1}{w^{m}}
\Big{(}\frac{f^2(w)}{f^2(z)}+2\frac{f(z)}{f(w)}\Big{)}
\Big{(}\frac{1}{(1+zw)^2}+\frac{1}{(z-w)^2}\Big{)} \0\\
\bar{U}_{nm}\!&=&\!\frac{-1}{3\sqrt{nm}}\oint\frac{dz}{2\pi i}
\oint\frac{dw}{2\pi i}
\frac{1}{z^{n}}\frac{1}{w^{m}} \Big{(}\frac{f^2(z)}{f^2(w)}+
2\frac{f(w)}{f(z)}\Big{)}
\Big{(}\frac{1}{(1+zw)^2}+\frac{1}{(z-w)^2}\Big{)} \0
\eea
By changing $z\to -z$ and $w\to -w$, it is easy to show that
\beq
(-1)^n U_{nm} (-1)^m = \bar U_{nm}, \quad \quad {\rm or}
\quad\quad C U = \bar U C, \quad\quad C_{nm} = (-1)^n \delta_{nm}\label{CU}
\eeq
In the second part of this equation we have introduced a matrix notation
which we will use throughout the paper.

The integrals can be directly computed in terms of the Taylor coefficients
of $f$. The result is
\bea
E_{nm}&=&(-1)^n\delta_{nm}\label{Anm1}\\
U_{nm}&=&\frac{1}{3\sqrt{nm}}\sum_{l=1}^{m}l\Big{[}(-1)^n
B_{n-l}B_{m-l}+2b_{n-l}b_{m-l}(-1)^m\0\\
       &&-(-1)^{n+l} B_{n+l}B_{m-l}-2b_{n+l}b_{m-l}(-1)^{m+l}\Big{]}\label{Unm1}\\
\bar{U}_{nm}&=&(-1)^{n+m}U_{nm}\label{Ubarnm1}
\eea
where we have set
\bea
f(z)&=&\sum_{k=0}^{\infty}b_kz^k\nonumber\\
f^2(z)&=&\sum_{k=0}^{\infty}B_kz^k, \quad \quad {\rm i.e.}\quad\quad
B_k = \sum_{p=0}^k b_p b_{k-p} \label{an}
\eea
Eqs.(\ref{Anm1}, \ref{Unm1}, \ref{Ubarnm1}) are obtained by expanding the
relevant integrands in powers of $z,w$ and correspond to the pole
contributions around the origin. We notice that the above integrands
have poles also
outside the origin, but these poles either are not in the vicinity of the
origin of the $z$ and $w$ plane, or, like the poles at $z=w$, simply give
vanishing contributions.

One can use this representation for (\ref{Unm1}, \ref{Ubarnm1}) to make
computer calculations. For instance it is easy to show that
the equations
\beq\label{v1}
\sum_{k=1}^{\infty} U_{nk}U_{km}=\delta_{nm},\quad\quad
\sum_{k=1}^{\infty} \bar U_{nk} \bar U_{km}=\delta_{nm}\label{UU}
\eeq
are satisfied to any desired order of approximation. Each identity follows
from the other by using (\ref{CU}). It is also easy to
make the identification
\beq
V_{nm}^{ab} = (-1)^{n+m}\sqrt{nm}\, N_{nm}^{ab}\label{identif1}
\eeq
of the Neumann coefficients with the standard three strings vertex
coefficients\footnote{The factor of $(-1)^{n+m}$ in (\ref{identif1})
arises from the fact that the original definition of the Neumann
coefficients (\ref{neumann}) in \cite{leclair1}
refers to the bra three strings vertex $\langle V_3|$, rather than
to the ket vertex like in (\ref{V3}); therefore the two definitions
differ by a $bpz$ operation.}. Using (\ref{UU}), together with the
decomposition (\ref{decomp}), it is easy to establish the commutativity
relation (written in matrix notation)
\beq
[CV^{ab}, CV^{a'b'}] =0 \label{commut}
\eeq
for any $a,b,a',b'$. This relation is fundamental for the next developments.

An analytic proof of eq.(\ref{UU}) is given in Appendix.

\subsection{Zero modes}

The Neumann coefficients involving one zero mode are given by \beq
N_{0m}^{ab}=-\frac{1}{m}\oint\frac{dw}{2\pi i}\frac{1}{w^{m}}f'_b(w)
\frac{1}{f_a(0)-f_b(w)}\label{N0} \eeq In this case too we make the decomposition \beq
N_{0m}^{ab}=\frac{1}{3}\Big{(}E_{m}+\bar{\a}^{a-b}U_{m}+\a^{a-b}
\bar{U}_{m}\Big{)}\label{N0decomp} \eeq where $E,U,\bar{U}$ can be given, after some algebra, the
explicit expression
\bea E_{n}&=&-\frac{4i}{n}\oint\frac{dw}{2\pi i}\frac{1}{w^{n}}
\frac{1}{1+w^2}\frac{f^3(w)}{1-f^3(w)}=\frac{2i^n}{n}\nonumber\\
U_{n}&=&-\frac{4i}{n}\oint\frac{dw}{2\pi i}
\frac{1}{w^{n}} \frac{1}{1+w^2}
\frac{f^2(w)}{1-f^3(w)} =\frac{\a_n}{n}\\
\bar{U}_{n}&=&(-1)^n\,U_{n}=(-1)^n\frac{\a_n}{n} \nonumber
\eea
The numbers $\a_n$ are Taylor coefficients
\bea
\sqrt{f(z)}&=&\sum_{0}^{\infty}\a_nz^n\nonumber
\eea
They are related to the $A_n$ coefficients of Appendix B of  \cite{RSZ2}
(see also \cite{GJ1}) as follows: $\a_n = A_n$ for $n$ even and $\a_n = i A_n$
for $n$ odd. $N_{0n}^{ab}$ are not related in a simple way as
(\ref{identif1}) to the corresponding three strings vertex coefficients.
The reason is that the latter satisfy the conditions
\beq
\sum_{a=1}^3 \, V_{0n}^{ab} =0\label{gaugefixing}
\eeq
These constraints fix the invariance $V_{0n}^{ab} \to V_{0n}^{ab} +
B_n^b$, where $B_n^b$ are arbitrary numbers, an invariance which arises
in the vertex (\ref{V3}) due to momentum conservation.
For the Neumann coefficients $N_{0n}^{ab}$ we have instead
\beq
\sum_{a=1}^3 \, V_{0n}^{ab}= E_n\0
\eeq
It is thus natural to define
\beq
\hat N_{0n}^{ab} =N_{0n}^{ab} - \frac 13 E_n\label{N0tilde}
\eeq
Now one can easily verify that\footnote{The $\sqrt{2}$ factor is
there because in \cite{RSZ2} the $\alpha'=1$ convention is used}
\beq
V_{0n}^{ab} = -\sqrt{2n}\, \hat N_{0n}^{ab}\label{identif2}
\eeq
It is somewhat surprising that in this relation we do not meet
the factor $(-1)^n$, which we would expect on the basis of the $bpz$
conjugation (see footnote after eq.(\ref{identif1})). However
eq.(\ref{identif2}) is also naturally requested by the integrable
structure found in \cite{BS}. The absence of the $(-1)^n$ factor
corresponds to the exchange $V_{0n}^{12} \leftrightarrow V_{0n}^{21}$.
This exchange does not seem to affect in any significant way the results
obtained so far in this field.

To complete the discussion about the matter sector one should recall
that beside eq.(\ref{UU}), there are other basic equations from which all
the results about the Neumann coefficients can be derived. They concern
the quantities
\bea
 W_n = - \sqrt {2n}\, U_n = - \sqrt {\frac 2n}\, \alpha_n,\quad\quad
 W_n^* = - \sqrt {2n}\, \bar U_n = - \sqrt {\frac 2n} \,(-1)^n\alpha_n
\label{Wn}
\eea 
The relevant identities, \cite{GJ1,RSZ2}, are
\beq
\sum_{n=1}^\infty W_n\, U_{np} = W_p, \quad\quad \sum _{n\geq 1} W_nW_n^* =
2V_{00}^{aa}\label{Wid}
\eeq
These identities can easily be shown numerically to be correct at any 
desired approximation. An analytic proof can presumably be obtained 
with the same methods as in Appendix.    

Finally let us concentrate on the Neumann coefficients $N_{00}^{ab}$.
Although a formula for them can be found in \cite{leclair1}, 
these numbers are completely arbitrary due to momentum conservation.
The choice
\beq
V_{00}^{ab} = \delta_{ab}\,\ln \frac {27}{16} \label{V00}
\eeq
is the same as in \cite{GJ1}, but it is also motivated by  
one of the most surprising and mysterious aspects of SFT, namely 
its underlying integrable structure: the matter Neumann
coefficients obey the Hirota equations of the dispersionless Toda
lattice hierarchy. This was explained in \cite{BS} following
a suggestion of \cite{BR}. On the basis of these equations
the matter Neumann coefficients with nonzero labels can be expressed
in terms of the remaining ones. The choice of (\ref{V00}) in this context is
natural.

\section{Ghost three strings vertex and $bc$ Neumann coefficients}

The three strings vertex for the ghost part is more complicated
than the matter part due to the zero modes of the $c$ field.
As we will see, the latter generate an ambiguity in the definition of the
Neumann coefficients. Such an ambiguity can however be exploited to
formulate and solve in a compact form the problem of finding solutions
to eq.(\ref{EOMg})\footnote{An alternative treatment of the ghost three--strings vertex has been given recently in \cite{Erler,Barso}} .

\subsection{Neumann coefficients: definitions and properties}

To start with we define, in the ghost sector, the vacuum states
$|\hat 0\rangle$ and $|\dot 0\rangle$ as follows
\beq
|\hat 0\rangle = c_0c_1|0\rangle, \quad\quad |\dot 0\rangle =
c_1|0\rangle
\label{vacua}
\eeq
where $|0\rangle$ is the usual $SL(2,\RR)$ invariant vacuum.
Using $bpz$ conjugation
\bea
c_n\rightarrow (-1)^{n+1}c_{-n},\quad\quad
b_n\rightarrow (-1)^{n-2}b_{-n}, \quad\quad |0\rangle \rightarrow \langle 0|
\eea
one can define conjugate states. It is important that, when applied to 
products of oscillators, the $bpz$ conjugation does not change the order
of the factors, but transforms rigidly the vertex and all the squeezed states we will consider in the sequel 
(see for instance  eq.(\ref{V3gh'}) below). 

The three strings interaction vertex is defined,
as usual, as a squeezed operator acting on three
copies of the $bc$ Hilbert space
\beq
\langle \tilde V_{3}|=\, _1\!\langle\hat{0}|\, _2\!\langle\hat{0}|\,
_3\!\langle\hat{0}|e^\tE,
\quad\quad
\tE=\sum_{a,b=1}^3\sum_{n,m}^{\infty}c_n^{(a)}\tN_{nm}^{ab}b_m^{(b)}
\label{V3gh}
\eeq
Under $bpz$ conjugation
\beq
|\tilde V_{3}\rangle=e^{\tE'}|\hat{0}\rangle_1|\hat{0}\rangle_2|\hat{0}
\rangle_3,\quad\quad
{\tE'}=-\sum_{a,b=1}^3\sum^\infty_{n,m}(-1)^{n+m}c_n^{(a)\,
\dagger}\tN_{nm}^{ab}b_m^{(b)\, \dagger}\label{V3gh'}
\eeq

In eqs.(\ref{V3gh}, \ref{V3gh'}) we have not specified the lower bound
of the $m,n$ summation. This point will be clarified below.\\
The Neumann coefficients $\tN_{nm}^{ab}$ are given by the contraction of
the $bc$ oscillators on the unit disk (constructed out of three
unit semidisks, as explained in section 3). They represent Fourier
components of the $SL(2,\RR)$ invariant $bc$ propagator (i.e. the propagator
in which the zero mode have been inserted at fixed points $\zeta_i$, $i =
1,2,3$):
\beq
\langle b(z) c(w)\rangle=\frac{1}{z-w}\prod_{i=1}^3
\frac{w-\zeta_i}{z-\zeta_i}
\eeq
Taking into account the conformal properties  of the $b, c$ fields
we get
\bea
\tN_{nm}^{ab}&=&\langle \tilde{V}_{123}|b_{-n}^{(a)}c_{-m}^{(b)}|\dot{0}\rangle_{123}\0\\
&=&\oint\frac{dz}{2\pi i}\oint\frac{dw}{2\pi i}\frac{1}{z^{n-1}}
\frac{1}{w^{m+2}}(f'_a(z))^2
\frac{-1}{f_a(z)-f_b(w)}\prod_{i=1}^3\frac{f_b(w)-\zeta_i}{f_a(z)-\zeta_i}
(f'_b(w))^{-1}\label{Nnmtil}
\eea

It is straightforward to check that
\beq
\tN_{nm}^{ab}=\tN_{nm}^{a+1,b+1}\label{cyclgh}
\eeq
and (by letting $z\rightarrow -z,\, w\rightarrow -w$)
\beq
\tN_{nm}^{ab}=(-1)^{n+m}\tN_{nm}^{ba}\label{twistgh}
\eeq
Now we choose  $\zeta_i=f_i(0)=\alpha^{2-i}$ so that the product factor
in (\ref{Nnmtil}) nicely simplifies as follows
\beq
\prod_{i=1}^3\frac{f_b(w)-f_i(0)}{f_a(z)-f_i(0)}=
\frac{f^3(w)-1}{f^3(z)-1}\, ,\, \quad\forall\,
a,b=1,2,3\0
\eeq
Now, as in the matter case, we consider the decomposition
\beq
\tN_{nm}^{ab}=\frac{1}{3}(\tE_{nm}+\bar{\alpha}^{a-b}\tU_{nm}+
\alpha^{a-b}\bar{\tU}_{nm})\label{decompgh}
\eeq
where
\bea
\tE_{nm}&=&\oint\frac{dz}{2\pi i}
\oint\frac{dw}{2\pi i}{\N}_{nm}(z,w){\A}(z,w)\nonumber\\
\tU_{nm}&=&\oint\frac{dz}{2\pi i}
\oint\frac{dw}{2\pi i}{\N}_{nm}(z,w){\U}(z,w)\\
\bar{\tU}_{nm}&=&\oint\frac{dz}{2\pi i}
\oint\frac{dw}{2\pi i}{\N}_{nm}(z,w)\bar{\U}(z,w)\nonumber
\eea
and
\bea
{\A}(z,w)\!&=&\!\frac{3f(z)f(w)}{f^3(z)-f^3(w)}\nonumber\\
{\U}(z,w)\!&=&\!\frac{3f^2(z)}{f^3(z)-f^3(w)}\nonumber\\
\bar{\U}(z,w)\!&=&\!\frac{3f^2(w)}{f^3(z)-f^3(w)}\0\\
{\N}_{nm}(z,w) \!&=&\! \frac 1{z^{n-1}}\frac 1{w^{m+2}}
(f'(z))^2 (f'(w))^{-1} \frac{f^3(w)-1}{f^3(z)-1}\0
\eea
After some elementary algebra, using
$f'(z)=\frac{4i}{3}\frac{1}{1+z^2}f(z)$, one finds
\bea
\tE_{nm}&=&\oint\frac{dz}{2\pi i}\oint\frac{dw}{2\pi i}\frac{1}{z^{n+1}}
\frac{1}{w^{m+1}}
\Big{(}\frac{1}{1+zw}-\frac{w}{w-z}\Big{)}\nonumber\\
\label{U}
\tU_{nm}&=&\oint\frac{dz}{2\pi i}\oint\frac{dw}{2\pi
i}\frac{1}{z^{n+1}}\frac{1}{w^{m+1}}\frac{f(z)}{f(w)} \Big{(}\frac{1}{1+zw}
-\frac{w}{w-z}\Big{)}\\
\bar{\tU}_{nm}&=&\oint\frac{dz}{2\pi i}\oint\frac{dw}{2\pi
i}\frac{1}{z^{n+1}}\frac{1}{w^{m+1}}\frac{f(w)}{f(z)} \Big{(}\frac{1}{1+zw}
-\frac{w}{w-z}\Big{)}\nonumber
\eea
Using the property $f(-z) = (f(z))^{-1}$, one can easily prove that
\beq
\bar{\tU}_{nm}  = (-1)^{n+m}{\tU}_{nm} \label{UUbar}
\eeq

\subsection{Computation of the coefficients}

In this section we explicitly compute the above integrals.
We shall see that the presence of the three $c$ zero modes
induces an ambiguity in the $(0,0)$, $(-1,1)$, $(1,-1)$ components
of the Neumann coefficients. This in turn arises
from the ambiguity in the radial ordering of the integration
variables $z,w$. While the result does not depend on what variable
we integrate first, it does depend in general
on whether $|z|>|w|$ or $|z|<|w|$.

If we choose $|z|>|w|$ we get
\beq
\tE_{nm}^{(1)}=\theta(n)\theta(m)(-1)^n\delta_{nm}+\delta_{n,0}\delta_{m,0}+
\delta_{n,-1}\delta_{m,1}
\eeq
while, if we choose $|z|<|w|$, we obtain
\beq
\tE_{nm}^{(2)}=\theta(n)\theta(m)(-1)^n\delta_{nm}-\delta_{n,1}\delta_{m,-1}
\eeq
where $\theta(n) = 1$ for $n>0$, $\theta(n) =0$ for $n\leq 0$.
We see that the result is ambiguous for the components $(0,0)$, $(-1,1)$, $(1,-1)$.

To compute $\tU_{nm}$ we expand $f(z)$ for small $z$, as in section 3,
\bea
f(z)&=&\sum_{k=0}^{\infty}b_kz^k\nonumber
\eea
Since $f^{-1}(z)=f(-z)$ we get the relation
\beq\label{as1}
\sum_{k=0}^{n}(-1)^kb_kb_{n-k}=\delta_{n,0}
\eeq
which is identically satisfied for $n$ odd, while for $n$ even it can be
also rewritten as
\beq\label{as2}
b_n^2=-2\sum_{k=1}^{n}(-1)^kb_{n-k}b_{n+k}
\eeq
Taking $|z|>|w|$ and integrating $z$ first, one gets\beq
\tU_{nm}^{(1a)}=\delta_{n+m}+(-1)^m\sum_{l=1}^n(b_{n-l}b_{m-l}-(-1)^l
b_{n-l}b_{m+l})\0
\eeq
If, instead, we integrate $w$ first,
\beq
\tU_{nm}^{(1b)}=(-1)^mb_nb_m+(-1)^m\sum_{l=1}^m(b_{n-l}b_{m-l}+
(-1)^l b_{n+l}b_{m-l})\0
\eeq
One can check that, due to (\ref{as2}),
\bea
\tU_{nm}^{(1a)}&=&\tU_{nm}^{(1b)}\equiv\tU_{nm}^{(1)}\label{U1}
\eea
Now we take $|z|<|w|$ and get similarly
\bea
&&\tU_{nm}^{(2a)}=(-1)^m\sum_{l=1}^n(b_{n-l}b_{m-l}-
(-1)^l b_{n-l}b_{m+l})\0\\
&&\tU_{nm}^{(2b)}=-\delta_{n+m}+(-1)^mb_nb_m+(-1)^m
\sum_{l=1}^m(b_{n-l}b_{m-l}
+(-1)^l b_{n+l}b_{m-l})\0
\eea
Again, due to (\ref{as2})
\bea
\tU_{nm}^{(2a)}&=&\tU_{nm}^{(2b)}=\tU_{nm}^{(2)}\label{U2}
\eea

Comparing $\tU^{(1)}$ with $\tU^{(2)}$, we see once more that the
ambiguity only concerns the $(0,0)$, $(-1,1)$, $(1,-1)$ components.
Using (\ref{decompgh}) we define
\bea
\tN_{nm}^{ab,\, (1,2)}&=&\frac{1}{3}(\tE_{nm}^{(1,2)}+\bar{\alpha}^{(a-b)}
\tU_{nm}^{(1,2)}+\alpha^{a-b}(-1)^{n+m}\tU_{nm}^{(1,2)})\nonumber
\eea
The above ambiguity propagates also to these coefficients, but only when
$a=b$. For later reference it is useful to notice that
\bea
&&\tN_{-1,m}^{ab,\, (1,2)} = 0, \quad {\rm except \quad perhaps\quad for}\quad
a=b,\quad m=1\0\\
&&\tN_{0,m}^{ab,\, (1,2)} = 0, \quad {\rm except \quad perhaps \quad for}\quad
a=b \quad m=0\label{zero}
\eea
and, for $|n|\leq 1$,
\beq
\tN_{n,1}^{ab,\, (1,2)} = 0, \quad {\rm except\quad perhaps 
\quad for}\quad
a=b\quad n=-1\0
\eeq

We notice that, if in eq.(\ref{V3gh},\ref{V3gh'}) the summation over $m,n$ starts
from $-1$, the above ambiguity is consistent with the general 
identification proposed in \cite{leclair1}
\beq
\tN^{ab}_{nm} = \langle \tilde V_3| b_{-n}^{(a)} c_{-m}^{(b)}|\dot{0}\rangle_1
|\dot{0}\rangle_2|\dot{0}\rangle_3 \label{NV}
\eeq
It is easy to see that the expression in the RHS is not $bpz$ covariant
when $(m,n)$ take values $(0,0)$, $(-1,1)$, $(1,-1)$ and the lower bound
of the $m,n$ summation in the vertex (see above) is $-1$. Such $bpz$
noncovariance corresponds exactly to the ambiguity we have come across
in the explicit evaluation of the Neumann coefficients. We can refer to
it as the {\it bpz} or {\it radial ordering anomaly}.

\subsection{Two alternatives}

It is clear that we are free to fix the ambiguity the way we wish,
provided the convention we choose is consistent with $bpz$ conjugation.
We consider here two possible choices.
The first consists in setting to zero all the components of the
Neumann coefficients which are ambiguous, i.e. the
$(0,0)$, $(-1,1)$, $(1,-1)$ ones.
This leads to a definition of the vertex (\ref{V3gh}) in which the
summation over $n$ starts from 1 while the summation over $m$ starts
from 0. In this way any ambiguity is eliminated and the Neumann
coefficients are $bpz$ covariant.
This is the preferred choice in the literature, \cite{GRSZ1,HK,HKw,
Oku1,Oku2}. In particular, it has led in \cite{GRSZ1} to a successful
comparison of the operator formulation with a twisted conformal field
theory one.

We would like, now, to make some comments about this first choice,
with the purpose of stressing the difference with the alternative one
we will discuss next. In particular we would like to emphasize some aspects of the BRST cohomology
in VSFT. In VSFT the BRST operator is conjectured \cite{GRSZ1, HKw} to take
the form
\beq
\Q=c_0+\sum_{n=1}^{\infty}f_n(c_n+(-1)^nc_{-n})\label{Q}
\eeq
It is easy to show that the vertex is BRST invariant, i.e.
\beq
\sum_{a=1}^3\Q^{(a)} |\tV_3\rangle =0\0
\eeq
Due to
\beq
\{\Q,b_0\}=1
\eeq
it follows that the cohomology of $\Q$ is trivial.
As was noted in \cite{Oku1}, this implies that the subset of the string
field algebra that solves (\ref{EOMg}) is the direct sum of
$\Q$--closed states and $b_0$--closed states (i.e. states in the Siegel gauge).
\beq
|\Psi\rangle=\Q|\lambda\rangle+b_0|\chi\rangle
\eeq
As a consequence of the BRST invariance of the vertex it follows
that the star product of a
BRST-exact state with any other is identically zero. This implies that
the VSFT equation of motion can determine only the Siegel gauge
part of the
solution.
\par
For this reason previous calculations were done with the use of
the \emph{reduced} vertex \cite{HKw, GRSZ1} which consists
of Neumann coefficients starting from the (1,1) component.
The unreduced star product can be recovered by the  midpoint insertion of $\Q=\frac{1}{2i}(c(i)-c(-i))$ as
\beq\label{redprod}
|\psi * \phi\rangle=\Q|\psi *_{b_0} \phi\rangle
\eeq
where $*_{b_0}$ is the reduced product.\par
In the alternative treatment given below, using an
enlarged Fock space, we compute the star product
and solve (\ref{EOMg}), without any gauge choice and any 
explicit midpoint insertion.

Motivated by the advantages it offers in the search of solutions 
to (\ref{EOMg}), we propose therefore
a second option. It consists in fixing the ambiguity by setting
\beq
\tN^{aa}_{-1,1} = \tN^{aa}_{1,-1} =0, \quad \quad \tN^{aa}_{0,0} =1.
\label{ambfix}
\eeq
If we do so we get a fundamental identity, valid for
$\tU_{nm} \equiv \tU_{nm}^{(1)}$
(for $n,m\geq 0$),
\beq
\sum_{k=0} \tU_{nk}\tU_{km} = \delta_{nm}\label{UU'}
\eeq
Defining
\beq
\tX^{ab} = C\tV^{ab},\0
\eeq
eq.(\ref{UU'}) entails
\beq
[\tX^{ab}, \tX^{a'b'}]=0\label{MM}
\eeq

One can prove eq.(\ref{UU'}) numerically. By using a cutoff in the
summation one can approximate the result to any desired order
(although the convergence with increasing cutoff is less rapid than
in the corresponding matter case, see section 3.1).
A direct analytic proof of eq.(\ref{UU'}) is given in Appendix.

The next subsection is devoted to working out some remarkable consequences
of eq.(\ref{UU'}).

\subsection{Matrix structure}

Once the convention (\ref{ambfix}) is chosen, we recognize that all the
matrices $(\tE,\tU,\bar{\tU})$ have the $(0,0)$ component equal to 1,
all the other entries of the upper row equal to 0, and a generally
non vanishing zeroth column.
More precisely
\bea
&&\tU_{00}=\tE_{00}=1\nonumber\\
&&\label{form} \tU_{n0}= b_{n}\quad \tE_{n0}=0,\quad \quad
\tU_{0n} = \tE_{0n}=\delta_{n,0} \\
&&\tU_{nm}\neq 0,\quad n,m>0\nonumber
\eea
This particular structure makes this kind of matrices simple to handle
under a generic analytic map $f$. In order to see this, let us
inaugurate a new notation, which we will
use in this and the next section. We recall that
the labels $M,N$ indicate the couple $(0,m), (0,n)$.  Given a matrix $M$,
let us distinguish between the `large' matrix $M_{MN}$ denoted
by the calligraphic symbol $\cal M$ and the `small' matrix $M_{mn}$
denoted by the plain symbol $M$.
Accordingly, we will denote by $\Y$  a matrix
of the form (\ref{form}), $\vec{y}= (y_1,y_2,...)$ will denote the
nonvanishing column vector and $Y$ the `small' matrix
\beq
\label{x} \Y_{NM}=\delta_{N0}\delta_{M0}+y_n\delta_{M0}+Y_{mn},
\eeq
or, symbolically, $\Y = (1, \vec{y},Y)$.

Then, using a formal Taylor expansion for $f$, one can show that
\beq\label{fx}
f[\Y]_{NM}=f[1]\delta_{N0}\delta_{M0}+{\Big (}\frac{f[1]-f[Y]}{1-Y}
\vec{y}{\Big )}_n\delta_{M0}+f[Y]_{mn}
\eeq

Now let us define
\bea
\Y &\equiv & \tX^{11}\nonumber\\
\Y_{+} &\equiv & \tX^{12}\label{X}\\
\Y_{-} &\equiv & \tX^{21}\label{Y+-}
\eea
These three matrices have the above form. Using (\ref{UU'}) one
can prove the following properties (which are well--known for the `small'
matrices)
\bea
&\Y+\Y_++\Y_-=1&\nonumber\\
&\Y^2+\Y_+^2+\Y_-^2=1&\nonumber\\
&\Y_+^3+\Y_-^3=2\Y^3-3\Y^2+1&\0\\
&\Y_+\Y_-=\Y^2-\Y&\label{bprop}\\
&[\Y,\Y_{\pm}]=0&\nonumber\\
&[\Y_+,\Y_-]=0&\nonumber
\eea
Using (\ref{x}, \ref{fx}) we immediately obtain (we point out that, in
particular for $\Y$, $y_{2n}= \frac 23 \,\,b_{2n}$, $y_{2n+1}=0$ and
$Y_{nm} = \tX_{nm}$
for $n,m>0$)
\bea
&Y+Y_++Y_-=1\nonumber\\
&\vec{y}+\vec{y}_++\vec{y}_-=0&\nonumber\\
&Y^2+Y_+^2+Y_-^2=1&\nonumber\\
&(1+Y)\vec{y}+Y_+\vec{y}_++Y_-\vec{y}_-=0&\0\\
&Y_+^3+Y_-^3=2Y^3-3Y^2+1&\0\\
&Y_+^2\vec{y}_+ +Y_-^2\vec{y}_- =(2Y^2-Y-1)\vec{y}&\label{lprop}\\
&Y_+Y_-=Y^2-Y&\0\\
&[Y,Y_{\pm}]=0&\nonumber\\
&[Y_+,Y_-]=0&\nonumber\\
&Y_+\vec{y}_-=Y\vec{y}= Y_-\vec{y}_+&\nonumber\\
&-Y_{\pm}\vec{y}=(1-Y)\vec{y}_{\pm}&\nonumber
\eea
These properties were shown in various papers, see \cite{HKw,Oku2}.
Here they are simply consequences of (\ref{bprop}), and therefore
of (\ref{UU'}). In particular we note
that the properties of the `large' matrices are isomorphic to
those of the `small' ones.
This fact allows us to work directly with the `large' matrices, handling
at the same time both zero and not zero modes.

\subsection{Enlarged Fock space}

We have seen in the last subsection the great advantages of introducing
the convention (\ref{ambfix}). In this subsection we make a proposal
as to how to incorporate
this convention in an enlargement of the $bc$ system's Fock space.
In fact, in order for eq.(\ref{NV}) to be consistent, a modification
in the RHS of this equation is in order. This can be done
by, so to speak, `blowing up' the zero mode sector. We therefore enlarge
the original Fock space, while warning that our procedure may be far from
unique. For each string, we split the modes $c_0$ and $b_0$:
\beq
\eta_0 \leftarrow c_0\rightarrow \eta_0^\dagger,\quad\quad
\xi_0^\dagger\leftarrow b_0\rightarrow \xi_0\0
\eeq
In other words we introduce two
additional couple of conjugate anticommuting creation and annihilation
operators $\eta_0, \eta_0^\dagger$ and $\xi_0,\xi_0^\dagger$
\beq
\{\xi_0, \eta_0\} = 1,\quad\quad
\{\xi_0^\dagger, \eta_0^\dagger\} = 1\label{xieta0}
\eeq
with the following rules on the vacuum
\bea
&&\xi_0|0\rangle =0,\quad\quad \langle 0|\xi_0^\dagger =0\label{xi0vac}\\
&&\eta_0^\dagger|0\rangle= 0,\quad\quad \langle 0|\eta_0=0\label{eta0vac}
\eea
while $\xi_0^\dagger, \eta_0$ acting on $|0\rangle$ create new states.
The $bpz$ conjugation properties are defined by
\beq
bpz(\eta_0) = -\eta_0^\dagger, \quad\quad bpz(\xi_0)= \xi_0^\dagger
\label{bpzxi0}
\eeq
The reason for this difference is that $\eta_0$ ($\xi_0$) is meant to
be of the same type as $c_0$ ($b_0$).
The anticommutation relation of $c_0$ and $b_0$ remain the same
\beq
\{c_0,b_0\} =1\label{b0c0}
\eeq
All the other anticommutators among these operators and with the other
$bc$ oscillators are required to vanish.
In the enlarged Fock space all the objects we have defined so far may get
slightly changed. In particular the three strings vertex
(\ref{V3gh},\ref{V3gh'}) is now defined by
\beq
\tE'_{(en)}=\sum^\infty_{n\geq1,m\geq 0}c_n^{(a)\,\dagger}
\tV_{nm}^{(ab)}b_m^{(b)\dagger}- \eta_0^{(a)}
b_0^{(a)}\label{V3gh''}
\eeq
With this redefinition of the vertex any ambiguity is eliminated,
as one can easily check.
In a similar way we may have to modify all the objects that enter into
the game.

The purpose of the Fock space enlargement is to make us able to 
evaluate vev's of the type
\beq
\langle\dot{0}|\exp \Big({cFb+c\mu+\lambda b}\Big)\,\exp\Big({c^{\dagger}Gb^{\dagger}
+\theta b^{\dagger}+c^{\dagger}\zeta}\Big)|\hat{0}\rangle\0
\eeq
which are needed in star products. Here
we use an obvious compact notation: $F,G$ denotes matrices
$F_{NM}, G_{NM}$, and $\lambda,\mu,\theta,\zeta$ are anticommuting
vectors $\lambda_N,\mu_N,\theta_N,\zeta_N$. In $cFb+c\mu+\lambda b$
it is understood that the mode
$b_0$ is replaced by $\xi_0$ and in
$c^{\dagger}Gb^{\dagger}+\theta b^{\dagger}+c^{\dagger}\zeta$ the mode
$c_0$ is replaced by $\eta_0$.  In this way the formula is
unambiguous and we obtain
\bea
&\langle\dot{0}|\exp\left({cFb+c\mu+\lambda b}\right)\exp\left({c^{\dagger}Gb^{\dagger}
+\theta b^{\dagger}+c^{\dagger}\zeta}\right)|\hat{0}\rangle&\nonumber\\
&& \0 \\
&=\det(1+FG)\,\exp\left({-\theta\frac{1}{1+FG}F\zeta-\lambda\frac{1}{1+
GF}G\mu-\theta\frac{1}{1+FG}\mu+\lambda\frac{1}{1+GF}\zeta}\right)&
\label{formula}
\eea
Eventually, after performing the star products, we will return 
to the original Fock space.

\section{Solving the ghost equation of motion in VSFT}

We are now ready to deal with the problem of finding a solution
to (\ref{EOMg})
\beq\label{ge}
\Q|\psi\rangle+|\psi\rangle\ast|\psi\rangle=0
\eeq
Since now we are operating in an enlarged the Fock space,
$\Q$ must be modified, with respect to the conjectured form of the BRST
operator (\ref{Q}) in VSFT, in the following way
\beq
\Q \to \Q_{(en)} = c_0-\eta_0 +\eta_0^\dagger
+\sum_{n=1}^{\infty}f_n(c_n+(-1)^nc_{-n})  \label{Qen}
\eeq
The first thing we would like to check is BRST invariance of the vertex, i.e.
\bea
\sum_{a=1}^{3}\Q_{(en)}^{(a)}|\tilde V_{3}\rangle_{(en)} =0\label{BRST}
\eea
It is easy to verify that both equations are identically satisfied
thanks to the first two eqs.(\ref{lprop}), and thanks to addition of
$-\eta_0$ in (\ref{Qen}) ($\eta_0^\dagger$ passes through and annihilates
the vacuum).

In order to solve equation (\ref{ge}) we proceed to find a
solution to
\beq
|\hat{\psi}\rangle_3
=_1\langle\dot{\psi}|_2\langle\dot{\psi}|V_{123}\rangle
\label{projecteq}
\eeq
where $\hat{\psi}$ and $\dot{\psi}$ are the same state on
the ghost number 2 and 1 vacuum, respectively. We choose
the following ansatz
\bea
|\hat{\psi}\rangle &=& |\hat{S}_{(en)}\rangle =\N
\exp\Bigg({\sum_{n,m\geq 1}
c_n^{\dagger}S_{nm}b_{m}^{\dagger}+\sum_{N\geq 0}c_N^{\dagger}
S_{N0}\xi_0^\dagger}\Bigg)|\hat{0}\rangle\label{Senhat}\\
|\dot{\psi}\rangle &=&|\dot{S}_{(en)}\rangle =\N
\exp\Bigg({\sum_{n,m\geq 1}c_n^{\dagger}S_{nm}b_{m}^{\dagger}+
\sum_{N\geq 0}c_N^{\dagger}S_{N0}\xi_0^\dagger}\Bigg)|\dot{0}\rangle\label{Sendot}
\eea

Following now the standard procedure, \cite{RSZ1, KP}, from
(\ref{projecteq}), using (\ref{formula}), we get
\bea
\T= \Y +(\Y_+,\Y_-)\frac 1{ 1-\Sigma{{\cal V}}}\Sigma
\left(\matrix{\Y_-\cr \Y_+}\right)\label{S2}
\eea
In RHS of these equations
\bea
\Sigma= \left(\matrix{\T&0\cr 0& \T}\right),
\quad\quad\quad
{\cal V} =
\left(\matrix{\Y&\Y_+\cr \Y_-&\Y}\right).\0
\eea
where $\T=C\cS$ and $\Y,\Y_\pm$ have been defined by eq.(\ref{Y+-}).

We repeat once more that the matrix equation (\ref{S2}) is understood
for `large' matrices, which include the zeroth row and column, i.e.
$\Y=\tX^{11}=C\tN^{11}= (1,\vec{y},Y\equiv\tX)$, $\T =(1,\vec{t}, \tT)$ and $\cS = (1,\vec{s},
\tS)$. This is a novelty of our treatment.
In fact, solving eq.(\ref{S2}), we obtain the algebraic equation
\beq
\T=C\cS=\frac{1}{2\Y}{\Big{(}}1+\Y-\sqrt{(1-\Y)(1+3\Y)}{\Big{)}}
\eeq
which splits into the relations
\bea
\label{sli}
\T_{00}&=&\cS_{00}=1\nonumber\\
\tT&=&\frac{1}{2\tX}{\Big{(}}1+\tX-\sqrt{(1-\tX)(1+3\tX)}{\Big{)}}\\
\vec{t}&=&\frac{1-\tT}{1-\tX}\vec{y}\nonumber
\eea
The normalization constant $\N$ is, formally, given by
\beq
\N = \frac 1{\det\left(1 -\Sigma {\cal V}\right)}\0
\eeq
However we notice that the (0,0) entry of $\Sigma\V$ is 1, so 
the determinant vanishes. Therefore we have to introduce a regulator
$\varepsilon \to 0$, and write
\beq
\N_\varepsilon = \frac 1\varepsilon \frac 1{\det'
\left(1 -\Sigma {\cal V}\right)}\label{norm}
\eeq
where $\det'$ is the determinant of the `small' matrix part alone. 
This divergence is not present in the literature, \cite{GRSZ1,Oku2}.
It is in fact related to the 1 eigenvalue of $\T$ and $\Y$ in the twist
even sector (i.e. in the eigenspace of $C$ with eigenvalue 1). This is
therefore an additional divergence with respect to the usual one
due to the 1 eigenvalue of $\tX$ in the twist--odd sector (see below).
 
Now we prove that this solves (\ref{ge}). Indeed,
after some elementary algebra, we arrive at the expression
\bea
\Q_{(en)}|\dot{S}\rangle +|\hat{S} \rangle&=& {\Big{(}}-
c_n^{\dagger}\left[(\vec{s})_n-(C-\cS)_{nk}f_k\right]
+c_0-\eta_0{\Big{)}}|\dot S\rangle
\label{Qproj}
\eea
We would like to find $\vec{f}$ so that the expression in square brackets in
(\ref{Qproj}) vanishes. Using the last equation in (\ref{sli}) we see
that this is true provided
\beq
\vec{y}= (1- \tX) \vec{f}\label{yf}
\eeq
Now, by means of an explicit calculation, we verify that the solution to (\ref{yf})
is
\beq
f_n=\frac{1}{2}(1+(-1)^n)(-1)^{\frac{n}{2}}\label{f}
\eeq

For inserting in the RHS of (\ref{yf}) both (\ref{f}) and $\tX$ in the form
\bea
\tX&=&\frac{1}{3}(1+C\tU+\tU C)\nonumber
\eea
we see that the vanishing of $f_n$ for $n$ odd is consistent
since $\vec{y}$ has no odd  components, while for $n$ even we have
\bea
\label{fe1}
y_{2n}= \sum_{k=1}^{\infty}\,\frac{2}{3}(-1)^k{\Big{(}}\delta_{2n,2k}-
\tU_{2n,2k}
{\Big{)}}
\eea

The second sum is evaluated with the use of the integral representation
of $\U$ (\ref{U})
\bea
\sum_{k=1}^\infty \,(-1)^k \,\tU_{2n,2k}&=& \oint\frac{dz}{2\pi i}\oint\frac{dw}{2\pi
i}\frac{1}{z^{2n+1}}\sum_{k=1}^{\infty}(-1)^k\frac{1}{w^{2k+1}}
\frac{f(z)}{f(w)} \Big{(}\frac{1}{1+zw}-\frac{w}{w-z}\Big{)}\nonumber\\
&=&- \oint\frac{dz}{2\pi i}\oint\frac{dw}{2\pi i}
\frac{1}{z^{2n+1}}\frac{1}{w}\frac{1}{1+w^2}\frac{f(z)}{f(w)}
\Big{(}\frac{1}{1+zw}-\frac{w}{w-z}\Big{)}\label{calcul}\\
&=&- \oint\frac{dz}{2\pi i}\frac{1}{z^{2n+1}}f(z)\Big{(}1-
\frac{1}{f(z)}\frac{1}{1+z^2}\Big{)}\nonumber\\
&=&-b_{2n}+ \sum_{k=1}^{\infty}(-1)^k
\delta_{2n,2k}\nonumber
\eea
The $\delta$-piece cancels with the one in (\ref{fe1}), while
the remaining one is precisely $y_{2n}$.

The derivation in (\ref{calcul}) requires some comments. In passing from the
first to the second line we use $\sum_{k=1}^{\infty}(-1)^k\frac{1}{w^{2k+1}}
= - \frac 1w \frac 1{1+w^2}$, which converges for $|w|>1$. Therefore, in order
to make sense of the operation, we have to move the $w$ contour outside
the circle of radius one. This we can do provided we introduce a regulator
(see Appendix) to avoid the overlapping of the contour with the branch points of $f(w)$,
which are located at $w=\pm i$. With the help of a regulator we move them
far enough and eventually we will move them back to their original position.
Now we can fully rely on the integrand in the second line of (\ref{calcul}).
Next we start moving the $w$ contour back to its original position around the
origin. In so doing we meet two poles (those referring to the
$\frac 1{1+w^2}$ factor), but it is easy to see that their contribution
neatly vanishes due to the last factor in the integrand. The remaining
contributions come from the poles at $w=z$ and at $w=0$. Their evaluation
leads to the third line in (\ref{calcul}). The rest is obvious.

As a result of this calculation we find that eq.(\ref{Qproj}) becomes
\beq
\Q_{(en)}|\dot{S}_{(en)}\rangle +|\hat{S}_{(en)}\rangle= (c_0-\eta_0)
|\dot{S}_{(en)}\rangle\label{Qproj'}
\eeq

Finally, as a last step, we return to the original Fock space. 
A practical rule to do so is to drop all the double zero mode
terms in the exponentials\footnote{Which is equivalent to normal ordering
these terms. We thank A.Kling and S.Uhlmann for this suggestion.}  (such as, for instance, $c_0\xi_0^\dagger$) and 
to impose the condition $c_0-\eta_0=0$ on the states, i.e. by considering
all the states that differ by  $ c_0-\eta_0$ acting on some state as
equivalent. The same has to be done also for $b_0 -\xi_0^\dagger$ (paying
attention not to apply both constraints simultaneously, because they do not
commute). These rules are enough for our purposes.
In this context the RHS of eq.(\ref{Qproj'}) is in the same class as 0.

Let us collect the results. In the original Fock space
the three string vertex is defined by
\beq
\tE' = \sum^\infty_{n\geq 1,M\geq 0}c_n^{(a)\,\dagger}
\tV_{nM}^{(ab)}b_M^{(b)\dagger} \label{V3gh'''}
\eeq
eqs.(\ref{Senhat},\ref{Sendot}) becomes
\bea
&& |\hat{S}\rangle =\N \exp\Bigg({\sum_{n,m\geq 1}
c_n^{\dagger}S_{nm}b_{m}^{\dagger}+\sum_{n\geq 1}c_n^{\dagger}
S_{n0}b_0}\Bigg)|\hat{0}\rangle\label{Shat}\\
&&|\dot{S}\rangle =\N
\exp\Bigg({\sum_{n,m\geq 1}c_n^{\dagger}S_{nm}b_{m}^{\dagger}}
\Bigg)|\dot{0}\rangle\label{Sdot}
\eea
It is now easy to prove, as a check, that
\beq\label{ge'}
\Q|\dot S\rangle+|\hat S\rangle=0
\eeq
where
\beq
\Q = c_0 + \sum_{n=1}^\infty (-1)^n(c_{2n} + c_{-2n})\label{trueQ}
\eeq
The above computation proves in a very direct way that the BRST
operator is nothing but the midpoint insertion ( $z=i$ ) of the operator
$\frac{1}{2i}(c(z)-c(\bar{z}))$ \cite{GRSZ1}. A different proof of this
identification, which makes use of the continuous basis of the $*$--algebra
\cite{GRSZ2}, was given in \cite{Oku2}.

As an additional remark, we point out that the ghost action calculated 
in the enlarged and restricted Fock space are different, although
they are both divergent due to the normalization (\ref{norm}).

A final warning to the reader: the method of `large' matrices is extremely
powerful and leads in a very straightforward way to results
that are very laborious to be obtained by alternative methods; the 
incorporation of `large' matrices in the Fock space formalism, on the
other hand, is given here on an {\it ad hoc} basis and certainly 
needs some formal polishing.

\subsection{A comment on the eigenvalues of Neumann coefficients matrices}

The introduction of `large' matrices gives us the opportunity to clarify 
an important 
point concerning the eigenvalues of the (twisted) matrices of three strings
vertex coefficients. 
The diagonalization of the reduced $*$-product was performed in 
\cite{Oku2, belov, belov2, Erler}, using a remarkable relation between 
matter coefficients and ghost coefficients \cite{GJ2}. Here we will make 
some comments about the singular role played by the midpoint, which turns 
out to be quite transparent in the diagonal basis.
Our commuting vertex coefficients (including the $(0,0)$ component) 
are of the form
\bea
\Y&=&\left(\matrix{1& 0\cr
                   \vec{y}& Y}\right)\\
\Y_\pm &=&\left(\matrix{0& 0\cr
                   \vec{y}_\pm & Y_\pm}\right)
\eea
It is then  evident that the eigenvalues are
\bea
eig[\Y]&=& 1 \oplus eig[Y]\\
eig[\Y_\pm]&=& 0 \oplus eig[Y_\pm]
\eea
We can easily put them in a block diagonal form
\bea
\hat\Y&=&\left(\matrix{1& 0\cr
                   0& Y}\right)\\
\hat\Y_\pm &=&\left(\matrix{0& 0\cr
                   0 & Y_\pm}\right)
\eea
This is achieved by
\beq
\hat\Y_{(\pm)}={\cal Z}^{-1}{\Y}_{(\pm)}{\cal Z}
\eeq
The block-diagonalizing  matrix is
\bea\label{V}
{\cal Z}&=&\left(\matrix{1& 0\cr
                   \vec{f}& 1}\right)\\
{\cal Z}^{-1}&=&\left(\matrix{1& 0\cr
                   -\vec{f}& 1}\right)
\eea
where
\beq\label{f'}
\vec{f}=\frac{1}{1-Y}\vec{y}=-\frac{1}{Y_\pm}\vec{y}_\pm
\eeq
At first sight one might think that, since (\ref{f'}) has the well known solution 
$f_n=\frac{1}{2}(1+(-1)^n)i^n$, it cannot be the case that either $Y$ 
has eigenvalue $1$, or $Y_\pm$ have the eigenvalue $0$. However (\ref{f'})
is a statement in the twist--even sector of the Hilbert space of vectors
$\{v_n\}, n=1,2,...$. So we must conclude that small matrices do not have
singular eigenvalues in this sector. The twist--odd sector on the contrary
contains singular eigenvalues. This was noted in \cite{belov,Erler},
where the following spectrum for the `small' 
matrices ,$(Y,Y_\pm)$ was found 
\bea\label{erler}
y(\kappa)&=& \frac{1}{2\cosh x -1}\\
y_\pm (\kappa) &=&\frac{\cosh x \pm \sinh x -1}{2\cosh x -1}
\eea
where $x\equiv\frac{\pi \kappa}{2}$ and $\kappa$ is a continuous 
parameter in the range $(-\infty,\, \infty)$.

It was pointed out in \cite{belov,Erler}
that to any $\kappa\neq 0$ there correspond two eigenvectors of opposite
twist--parity, while $\kappa=0$ has only one twist--odd eigenvector.
On the basis of the discussion in this paper we see that, when
considering  the `large' matrix $\Y$, we have an additional 1 eigenvalue 
whose twist--even eigenvector is given by the first column of (\ref{V})
\footnote{This is very similar to what happens with the Neumann 
coefficients in the matter sector with zero modes in \cite{bofeng}.}
.

In terms of the $bc$ standard modes we can write 
\bea\label{midp}
\tilde c_0&=&c_0 +\sum_{n\geq 1}f_n(c_n+(-1)^nc_n^\dagger)=\Q\\
\tilde c_n&=&c_n\quad n\neq 0\\
\tilde b_0&=&b_0\\
\tilde b_n&=&-f_nb_0+b_n\quad n\neq 0
\eea
where we have defined $(f_{-n}\equiv f_n)$.
As  noted also in \cite{Oku1} this is an equivalent representation 
of the $bc$ system\footnote{In order to prove this, twist invarince 
of $\vec{f}$ is crucial ($C\vec{f}=\vec{f}$)}
\beq
\{\tilde b_N,\tilde c_M\}=\delta_{N+M}\quad N,M=-\infty,...,0,...,\infty
\eeq
This  can possibly be viewed as a redefinition of the CFT fields 
$c(z),b(z)$.\\
 
If we exclude the $\tilde c_0,\tilde b_0$ zero modes we are left with 
the reduced $*$-product which turns out to be an associative  
product for states in the Siegel gauge.
In particular we note that, since, in the reduced product, 
the value $\kappa=0$ is related to only one eigenvector of $Y$
and since the same value is intimately 
related to the midpoint (\ref{midp}), the Siegel gauge can be regarded
as a split string description.
\par

A possible suggestion to appropriately treat the additional
singular 1 eigenvalue of $\Y$  
may be to ``blow up'' the zero modes, as before, in two triples of 
conjugate operators. Then we can safely compute $*$-products in an enlarged Fock 
space and then return to the original space by appropriate restrictions.
As an encouraging indication in this direction, we notice that
\beq
K_1=L_1+L_{-1}
\eeq
where $L_1$ and $L_{-1}$ are the ghost Virasoro generators,
can be modified in the enlarged space by formally setting 
\beq
c_0\rightarrow c_0-\eta_0 +\eta_0^\dagger
\eeq
It is easy to see that the vertex is $K_1$-invariant. In fact,
keeping for simplicity 
only the zero mode part which is the only one of interest in this aspect, 
we have
\beq
K_1\e^{-\eta_0b_0}|\hat 0\rangle=
\e^{-\eta_0b_0}\left( (-c_0-\eta_0 
+\eta_0^\dagger +\eta_0)(b_1+b_{-1})\right)c_0|\dot 0\rangle=0
\eeq

\section{Matter projectors and D--branes}

In this second part of the review we will be concerned with
the solutions of the matter part of the VSFT equations of motion,
i.e. with solutions to (\ref{EOMm}). However instead of reviewing
the well--known squeezed state solutions to \cite{KP,RSZ2} as well as
the related solutions discussed in \cite{RSZ3,GT}, our leading idea
will be to find VSFT solutions in order to make a comparison  with
solutions in scalar noncommutative field theories. In particular we will show
that it is possible to establish a one-to-one correspondence
between tachyonic lumps, i.e lower dimensional D--branes, in the
former and solitonic solutions in the latter. This correspondence is
fully exposed by introducing a constant background $B$ field.

To start with we fix the solution of the ghost part in the form given
in the previous section and concentrate on the matter part.
The value of the action for such solutions is given by
\bea
{\cal S}(\Psi) = {\EuScript K} \langle\Psi_m|\Psi_m\rangle\label{action2}
\eea
where ${\EuScript K}$ contains the ghost contribution. We recall that
${\EuScript K}$  is infinite (see above) unless the action
is suitably regularized. The choice of a regularization should be
understood as a `gauge' freedom, \cite{RSZ3}, in choosing the
solutions to (\ref{EOM}). So 
a coupled solution to (\ref{EOMg}) and (\ref{EOMm}), even if the action
is naively infinite in its ghost component, is nevertheless a
legitimate representative of the corresponding class of solutions.

\subsection{ Solitons in noncommutative field theories}

Before we set out to discuss solutions to (\ref{EOMm})
and to better explain our aim, let us briefly describe
solitons in noncommutative field theories, (see \cite{Komaba} for a
beautiful review). Noncommutative field
theories are effective low energy field theories which live on
the world-volume
of D--branes in the presence of a constant background $B$ field.
To be definite, let us think of a D25--brane in bosonic string
theory.
The simplest example of effective theories is a noncommutative
theory of a scalar field $\phi$, which is thought to represent the
tachyon living on the brane (this is an oversimplified situation,
in fact one could easily take the gauge field as well into account,
while the massive string states are thought
to having been integrated out).
We concentrate on the case in which $B$ is switched on along two
space directions,
which we denote by $x=q$ and $y=p$ (from now one, for simplicity,
we will drop the other
coordinates). The coordinates become noncommutative
and the ensuing situation can be described by replacing the initial
theory with a theory in which all products are replaced by the
Moyal $\star$ product with deformation parameter $\theta$
(linked to $B$ as explained below).
Alternatively one can use the Weyl map and replace the
noncommutative coordinates
by two conjugate operators $\hat q$, $\hat p$, such that
$[\hat q,\hat p]= i \theta$.
In the large $\theta$ limit, after a suitable rescaling of the
coordinates,
the kinetic part of the action becomes negligible, so only the
potential part, $\int dx\,dy \, V_\star (\phi)$, is relevant.
Using the
Weyl correspondence, the action can be replaced by
$2\pi\theta \,{\rm Tr}_{\cal H} V(\hat \phi)$, where
${\cal H}$ is the Hilbert space constructed out of
$\hat q,\hat p$, and  $\hat \phi$
is the operator corresponding to the noncommutative field $\phi$.
Solutions to the equations of motion take the form
\bea
\hat \phi = \lambda_i P, \quad\quad P^2=P\label{P}
\eea
where $\lambda_i$, $i=1,\ldots,n$ are the minima of the classical commutative
potential $V$, which is assumed to be polynomial.
The energy of such a solution
is therefore given by $2\pi\theta \,V(\lambda_i)\,{\rm Tr}_{\cal H} P$.

On the basis of this discussion it is clear that, in order to
know the finite
energy solutions of the noncommutative scalar theory, we have to
find
the finite rank projectors in the space ${\cal H}$. The latter
can be constructed
in the following way.  Define the harmonic oscillator
$a = (\hat q +i \hat p)/\sqrt{(2\theta)}$ and its hermitean conjugate
$a^\dagger$:
$[a,a^\dagger]=1$. By a standard construction we can define the
normalized
harmonic oscillator eigenstates:
$|n\rangle = \frac {(a^\dagger)^n}{\sqrt n!}|0\rangle$.
Now, via the Weyl correspondence, we can map any operator
$|n\rangle\langle m|$
to a classical function of the coordinates $x,y$. In particular $|n\rangle\langle n|$,
which are rank one projectors, will be mapped to classical functions
\bea
\psi_n(x,y)=  2 \, (-1)^n L_n \bigg(\frac {2r^2}\theta\bigg)
e^{- \frac {r^2}\theta}\label{solitons}
\eea
where $r^2 = x^2+y^2$. Each of these solutions, by construction, satisfy
$\psi_n\star\psi_n =\psi_n$. We refer to them as the GMS solitons \cite{GMS}.
They can be interpreted as D23--branes.
We can of course consider any finite sum of these projectors.
They will
also be solutions. They are interpreted as collapsing D23--branes.
Moreover, using the shift operator
$S= \sum_{n=0}^\infty |n+1\rangle\langle n|$,
one can set up a solution generating techniques, whereby a nontrivial
soliton solution can be generated starting from a trivial one by repeated application
of $S$ (see (\cite{Komaba})).

Our purpose in this paper is to show how the soliton solutions just described
arise in VSFT. The way they can be seen is by taking
the low energy limit of tachyonic lumps representing D23--branes in VSFT \cite{David,Mukho}.
They are, so to speak, relics of VSFT branes in the $\alpha'\to 0$ limit.

\subsection{ The background $B$ field}

But before we turn to this, we will introduce a background field $B$
in SFT. In ref. \cite{W2} and \cite{Sch} it was shown
that, when such a field is switched on, in the low energy limit the string field theory
star product factorizes into the ordinary Witten $*$ product and
the Moyal $\star$ product. A related result can be obtained in the
following way. The string field theory action (\ref{sftaction})
can be explicitly calculated in terms of local fields, provided
the string field is expressed itself in terms of local fields
as in (\ref{stringfield}).
Of course this makes sense only in the limit in which string theory can be
approximated by a local field theory. In this framework
(\ref{sftaction}) takes the form of
an infinite series of integrated local polynomials
(kinetic and potential terms) of the fields involved in
(\ref{stringfield}) as explained in section 2.
Now, it has been proven by \cite{sugino,KT} that, when a $B$
field is switched on, the kinetic term of (\ref{sftaction}) remains
the same while the three string vertex changes, being multiplied by
a (cyclically invariant) noncommutative phase factor
(see (\cite{sugino,KT}) and eq.(\ref{EtoE'})  below). It is easy to
see on a  general basis that the overall effect of such
noncommutative factor is to replace the ordinary product with the
Moyal product in the RHS of the effective action (for a related
approach see \cite{MRVY}).

Therefore, we know pretty well the effects of a $B$ field when perturbative
configurations are involved.  What we wish to explore here are the effects
of a $B$ field on nonperturbative solutions. We find that a $B$ field has the virtue of
smoothing out some of the singularities that appear in VSFT. As for
the overall star product in the presence of a background $B$ field, it turns out
that Witten's star product and Moyal product are completely entangled in
nonperturbative configurations. Nevertheless, in the low energy limit,
we can again witness the factorization into Witten's star product and the Moyal one.
It is exactly this factorization that allows us to recover the noncommutative
field theory solitons.

We also remark that switching on a $B$ field in VSFT is consistent
with the interpretation of VSFT. The latter
is thought to describe closed string theory, and the antisymmetric
field $B$ belongs in the massless sector of a bosonic closed string theory.

Finally we would like to stress that the Moyal star product referred to here
has nothing to do with the Moyal representation of Witten's star product
which was suggested in \cite{bars,doug}. This representation can be
seen as a confirmation of an old theorem \cite{Lich} concerning the uniqueness
of the Moyal product in the class of noncommutative associative products, however it is
realized in an (unphysical) auxiliary space, see \cite{Bo2,Chen,belov},
therefore it cannot affect the physical space--time.

\subsection{ Organization of the paper. Second part}

In section 7 we derive the new Neumann
coefficients for the three string vertex in the presence of a background $B$ field.
In section 8 we solve the projector equation (\ref{EOMm}) for
a 23--dimensional tachyonic lump and justify its D23--brane
interpretation. In section 9 we start examining the effect of the $B$
field on such
solution. In section 10 we generalize the lump solution of section 8.
We construct a series
of solutions to the matter projector equation, which we denote by
$|\Lambda_n\rangle$ for any natural number $n$.
$|\Lambda_n\rangle$ is generated by acting on a tachyonic lump
solution $|\Lambda_0\rangle$ with $(-\kappa)^nL_n(\x/\kappa)$, where
$L_n$ is the $n$-th Laguerre polynomial,
$\x$ is a quadratic expression in the string creation
operators, see below eqs.(\ref{x}, \ref{Lambdan}), and $\kappa$ is an
arbitrary real
constant.  These states satisfy the remarkable properties
\bea
&& |\Lambda_n\rangle * |\Lambda_m\rangle = \delta_{n,m}
|\Lambda_n\rangle \label{nstarm}\\
&& \langle \Lambda_n |\Lambda_m\rangle = \delta_{n,m}
 \langle \Lambda_0 |\Lambda_0\rangle \label{nm}
\eea
Each $|\Lambda_n\rangle$ represents a D23--brane, parallel to all the
others. In section 11 we show that
the field theory limit of $|\Lambda_n\rangle$ factors into
the sliver state (D25--brane) and the $n$-th GMS soliton.
Section 12 describes related results.

\section{The three string vertex in the presence of a constant
background B field}

The three string vertex \cite{W1,GJ1,GJ2} of the Open String Field
Theory was given in eq.(\ref{V3}).
The notation $V_{MN}^{rs}$ for the vertex coefficients will often be used
from now on, where $M(N)$ will denote the couple
$\{0,m\}$ ($\{0,n\}$).

Our first goal is to find the new form of the coefficients $V_{MN}^{rs}$
when a constant $B$ field is switched on. We start from the simplest
case, i.e. when $B$ is nonvanishing in two space directions,
say the $24$--th and $25$--th ones.
Let us denote these directions with the Lorentz indices $\alpha$ and
$\beta$.
Then, as is well--known \cite{SW,sugino,KT}, in these two directions
we have a new effective metric $G_{\alpha\beta}$,
the open string metric, as well as
an effective antisymmetric parameter $\theta_{\alpha\beta}$, given by
\bea
G^{\alpha\beta} = \left(\frac {1}{\eta + 2\pi \alpha' B}\,\eta\,
\frac {1}{\eta - 2\pi \alpha' B}\right)^{\alpha\beta},\quad\quad
\theta^{\alpha\beta} = -(2\pi \a')^2\left(\frac {1}{\eta +
2\pi \alpha' B}\,B\,
\frac {1}{\eta - 2\pi \alpha' B}\right)^{\alpha\beta}\0
\eea
Henceforth we set $\a' =1$, unless otherwise specified.

The three string vertex is modified only in the 24-th and
25-th direction, which, in view of the subsequent D--brane
interpretation, we call the transverse directions.
We split the three string vertex into the tensor product of the
perpendicular part and the parallel part
\bea
|V_3\rangle = |V_{3,\perp}\rangle \, \otimes\,|V_{3,_\|}\rangle
\label{split}
\eea
The parallel part is the same as in the ordinary case and will not be
re-discussed here. On the contrary we will describe in detail the
perpendicular part of the vertex. We rewrite the exponent $E$ as
$E=E_\|+ E_\perp$, according to the above splitting.
$E_\perp$ will be modified as follows
\bea
E_\perp&\to& E'_\perp = \sum_{r,s=1}^3\left(\frac 12 \sum_{m,n\geq 1}
G_{\alpha\beta}a_m^{(r)\alpha\dagger}V_{mn}^{rs}
a_n^{(s)\beta\dagger} +
\sum_{n\geq 1}G_{\alpha\beta}p_{(r)}^{\alpha}V_{0n}^{rs}
a_n^{(s)\beta\dagger}\right.\0\\
&&\left. \quad\quad +\,\frac 12 G_{\alpha\beta}p_{(r)}^{\alpha}
V_{00}^{rs}p_{(s)}^\beta+
\frac i2 \sum_{r<s} p_\alpha^{(r)}\theta^{\alpha\beta}
p_\beta^{(s)}\right)\label{EtoE'}
\eea

Next, as far as the zero modes are concerned, we pass from the
momentum to the oscillator basis, \cite{GJ1,GJ2}. We define
\bea
a_0^{(r)\alpha} = \frac 12 \sqrt b \hat p^{(r)\alpha}
- i\frac {1}{\sqrt b} \hat x^{(r)\alpha},
\quad\quad
a_0^{(r)\alpha\dagger} = \frac 12 \sqrt b \hat p^{(r)\alpha} +
i\frac {1}{\sqrt b}\hat x^{(r)\alpha}, \label{osc}
\eea
where $\hat p^{(r)\alpha}, \hat x^{(r)\alpha}$ are the zero momentum
and position operator of the $r$--th string, and we have kept the
`gauge' parameter $b$ of ref.\cite{RSZ2} ($b\sim {\a'}$). From now on
Lorentz indices are raised and lowered by means of the effective
open string metric, for instance $p^{(r)\alpha} = G^{\a\b}p^{(r)}_\b$.
We have
\bea
\big[a_M^{(r)\alpha},a_N^{(s)\beta\dagger}\big]=
G^{\alpha\beta}\delta^{rs}\delta_{MN},\quad\quad\quad N,M\geq 0
\label{a0a0}
\eea
Denoting by $|\Omega_{b,\theta}\rangle$ the oscillator vacuum
(\,$a_N^\alpha|\Omega_{b,\theta}\rangle=0$, for $N\geq 0$\,),
the relation
between the momentum basis and the oscillator basis is defined by
\bea
&&|p^{24}\rangle_{123}\otimes|p^{25}\rangle_{123}\equiv
|\{p^\alpha\}\rangle_{123}
=\0\\
&& \nonumber \\
&&\left(\frac b{2\pi\sqrt {{\rm det}G}}\right)^\frac 32 {\rm exp}
\left[\sum_{r=1}^3
\left(- \frac b4 p^{(r)}_\alpha G^{\alpha\beta}p^{(r)}_\beta+
\sqrt b  a_0^{(r)\alpha\dagger}p^{(r)}_\a
- \frac 12 a_0^{(r)\alpha\dagger}G_{\alpha\beta}a_0^{(r)\beta\dagger}
\right)\right]|\Omega_{b,\theta}\rangle\0
\eea

Now we insert this equation inside $E'_\perp$ and try to eliminate the
momenta along the perpendicular directions by integrating them out.
To this end we rewrite $E'_\perp$ in the following way and, for
simplicity, drop all the labels $\alpha,\beta$ and $r,s$:
\bea
E'_\perp = \frac 12\sum_{m,n\geq 1}a_m^\dagger GV_{mn}a_n^\dagger +
\sum_{n\geq 1}
pV_{0n}a_n^\dagger + \frac 12 p\left[G^{-1}(V_{00}+\frac b2) +
\frac i2 \theta \epsilon \chi \right]p
-\sqrt b p  a_0^\dagger +
\frac 12 a_0^\dagger G a_0^\dagger\0
\eea
where we have set $\theta^{\alpha\beta}=
\epsilon^{\alpha\beta}\theta$ and introduced
the matrices $\epsilon$ with entries
$\epsilon^{\alpha\beta}$ (which represent the $2\times 2$ antisymmetric
symbol with $\epsilon^1{}_2=1$) and $\chi$ with entries
\bea
\chi^{rs}= \left(\matrix{0&1&-1\cr -1&0&1\cr 1&-1&0}\right)\label{chi}
\eea
At this point we impose momentum conservation. There are three
distinct ways to do that and eventually one has to (multiplicatively)
symmetrize with respect to them. Let us start by setting
$p_3=-p_1-p_2$ in $E'_\perp$ and obtain an expression of the form
\bea
p\, X_{00}\, p + \sum_{N\geq 0} p\, Y_{0N}\, a_N^\dagger+
\sum_{M,N\geq 0}a_M^\dagger\, Z_{MN }\, a_N^\dagger\label{pXp}
\eea
where, in particular, $X_{00}$ is given by
\bea
X_{00}^{\alpha\beta,rs}= G^{\alpha\beta}\, (V_{00}+ \frac b2) \,
\eta^{rs}+ i \frac \theta {4} \,\epsilon^{\a\b}\, \epsilon^{rs}
\label{X00}
\eea
Here the indices $r,s$ take only the values 1,2, and
$ \eta =$ \mbox{ \small $ \left(\matrix{1 & 1/2\cr 1/2 & 1}\right)$}.

Now, as usual, we redefine $p$ so as eliminate the linear term in
(\ref{pXp}). At this point we can easily perform the Gaussian
integration over $p_{(1)},p_{(2)}$, while the remnant of
(\ref{pXp}) will be expressed in terms of the inverse of
$X_{00}$:
\bea
\left(X_{00}^{-1}\right)^{\a\b,rs}=
\frac {2A^{-1}}{4a^2+3}\left(\frac 32\,  G^{\a\b}\,
(\eta^{-1})^{rs} -2i \,a\, \epsilon^{\a\b} \,
\epsilon^{rs}\right)\label{X00-1}
\eea
where
\bea
A = V_{00}+ \frac b2,  \quad\quad\quad a =
\frac \theta{4A }.
\label{definitions}
\eea
Let us use henceforth for the $B$ field the explicit form
\bea
 B_{\a\b}= \left(\matrix {0&B\cr -B&0\cr}\right)
\label{B}
\eea
so that
\bea
{\rm Det G} = \left( 1+ (2 \pi B)^2\right)^2, \quad \quad
\theta  = - (2\pi )^2 B,
\quad\quad a= -\frac {\pi^2}A\, B\label{thetaB}
\eea

Now one has to symmetrize with respect to the three possibilities
of imposing the momentum conservation. Remembering the factors due
to integration over the momenta and collecting the results one
gets for the three string vertex in the presence of a $B$ field
\bea
|V_3 \rangle' = |V_{3,\perp}\rangle ' \,\otimes\,|V_{3,\|}\rangle
\label{split'}
\eea
$|V_{3,\|}\rangle$ is the same as in the ordinary case
(without $B$ field), while
\bea
|V_{3,\perp}\rangle'= K_2\, e^{-E'}|\tilde 0\rangle\label{V3'}
\eea
with
\bea
&&K_2= \frac {\sqrt{2\pi b^3}}{A^2 (4a^2+3)}({\rm Det} G)^{1/4},
\label{K2}\\
&&E'= \frac 12 \sum_{r,s=1}^3 \sum_{M,N\geq 0} a_M^{(r)\a\dagger}
\EV_{\a\b,MN}^{rs} a_N^{(s)\b\dagger}\label{E'}
\eea
and $|\tilde 0\rangle = |0\rangle \otimes |\Omega_{b,\theta}\rangle$.
The coefficients $\EV_{MN}^{\a\b,rs}$ are given by
\bea
&&\EV_{00}^{\a\b,rs} = G^{\a\b}\delta^{rs}- \frac {2A^{-1}b}{4a^2+3}
\left(G^{\a\b} \phi^{rs} -ia \epsilon^{\a\b}\chi^{rs}\right)
\label{VV00}\\
&&\EV_{0n}^{\a\b,rs} = \frac {2A^{-1}\sqrt b}{4a^2+3}\sum_{t=1}^3
\left(G^{\a\b} \phi^{rt} -ia \epsilon^{\a\b}\chi^{rt}\right)
V_{0n}^{ts}\label{VV0n}\\
&&\EV_{mn}^{\a\b,rs} = G^{\a\b}V_{mn}^{rs}-
\frac {2A^{-1}}{4a^2+3}\sum_{t,v=1}^3
V_{m0}^{rv}\left(G^{\a\b} \phi^{vt}
-ia \epsilon^{\a\b}\chi^{vt}\right)V_{0n}^{ts}\label{VVmn}
\eea
where, by definition, $V_{0n}^{rs}=V_{n0}^{sr}$, and
\bea
\phi= \left(\matrix{1& -1/2& -1/2\cr
                    -1/2& 1& -1/2\cr
                     -1/2 &-1/2 &1}\right)\label{phi}
\eea
while the matrix $\chi$ has been defined above (\ref{chi}). These
two matrices satisfy the algebra
\bea
\chi^2 = - 2\phi,\quad\quad \phi\chi=\chi\phi = \frac 32 \chi,
\quad\quad \phi^2= \frac 32 \phi\label{chiphi}
\eea

Next, let us notice that the above results
can be easily extended to the case in which the transverse directions
are more than two (i.e. the 24--th and 25--th ones) and even. The
canonical form of the transverse $B$ field is
\bea
B_{\a\b} = \left(\matrix{\matrix{0&B_1\cr -B_1&0\cr} & 0& \ldots\cr
0& \matrix{0&B_2\cr -B_2&0\cr}&\ldots\cr
\ldots& \ldots& \ldots\cr}\right) \label{genB}
\eea
It is not hard to see that each couple of conjugate transverse
directions under this decomposition, can be treated in a completely
independent way. The result is that each couple of directions
$(26-i,25-i)$, corresponding to the eigenvalue $B_i$, will be
characterized by the same formulas (\ref{VV00}, \ref{VV0n}, \ref{VVmn})
above with $B$ replaced by $B_i$.

The properties of the new Neumann coefficients $\EV_{NM}^{rs}$ have been
analyzed in \cite{BMS1}. Here we write down the results.

To start with, let us quote
\begin{itemize}
\item (i) $\EV_{NM}^{\a\b,rs}$ are symmetric under simultaneous exchange of
the three couples of indices;
\item (ii) they are endowed with the property of cyclicity in the
$r,s$ indices, i.e. $\EV^{rs}= \EV^{r+1,s+1}$, where $r,s=4$ is
identified with $r,s=1$.
\end{itemize}

Next let us extend the twist matrix $C$ by $C_{MN}= (-1)^M
\delta_{MN}$ and define
\bea
\EX^{rs} \equiv C \EV^{rs}, \quad\quad r,s=1,2,
\quad\quad \EX^{11}\equiv \EX\label{EX}
\eea
These matrices commute
\bea
[\EX^{rs}, \EX^{r's'}] =0 \label{commute}
\eea
and
\bea
(\EX^{rs})^* = \tilde \EX^{rs}, \quad {\rm i.e.}\quad (\EX^{rs})^\dagger=
\EX^{rs}\0
\eea
Moreover we have the following properties, which mark a difference
with the $B=0$ case,
\bea
C\EV^{rs}= \tilde \EV^{sr}C ,\quad\quad C\EX^{rs}= \tilde \EX^{sr}C
\eea
where we recall that tilde denotes transposition with respect to the
$\a,\b$ indices. Finally one can prove that
\bea
&&\EX^{11}+ \EX^{12}+ \EX^{21} = \I\0\\
&& \EX^{12}\EX^{21} = (\EX^{11})^2-\EX\0\\
&& (\EX^{12})^2+ (\EX^{21})^2= \I- (\EX^{11})^2\0\\
&& (\EX^{12})^3+ (\EX^{21})^3 = 2 (\EX^{11})^3 -  3(\EX^{11})^2 +\I
\label{Xpower}
\eea
In the matrix products of these identities, as well as throughout
the paper, the indices $\a,\b$ must be understood in alternating
up/down position: $\EX^{\a}{}_{\b}$. For instance, in (\ref{Xpower})
$\I$ stands for $\delta^\a{}_\b\,\delta_{MN}$.

\section{The squeezed state solution}

In this section we wish to find a solution to the equation of motion
$|\Psi\rangle * |\Psi\rangle = |\Psi\rangle$ in the form of squeezed
states \cite{okuda,FKM,GRSZ2,Sch1}.
A squeezed state in the present context is written as
\bea
|S\rangle = |S_\perp\rangle\,\otimes\,|S_\|\rangle\label{factor}
\eea
where $|S_\|\rangle$ has the ordinary form, see \cite{KP,RSZ2}, and is
treated in the usual way, while
\bea
|S_\perp\rangle \,=\, {\cal N}^2 \,{\rm exp}\left(-\frac 12
\sum_{M,N\geq 0}
a_M^{\a\dagger}\,\ES_{\a\b,MN}\, a_N^{\b\dagger}\right)
|\Omega_{b,\theta}\rangle
\label{squeezed1}
\eea
The $*$ product of  two such states, labeled $_1$ and $_2$, is
\bea
|S'_\perp\rangle =|S_{1,\perp}\rangle\,*\,|S_{2,\perp}\rangle=
\frac{K_2\,({\cal N}_1{\cal N}_2)^2}{{\rm DET}({\bf I}-
\Sigma{\cal V})^{1/2}}\, \,
{\rm exp}\left(-\frac 12 \sum_{M,N\geq 0}
a_M^{\a\dagger}\ES'_{\a\b,{MN}}a_N^{\b\dagger}\right)
|\tilde 0\rangle \label{squeez}
\eea
where, in matrix notation which includes both the indices $N,M$
and $\a,\b$,
\bea
\ES'= \EV^{11} +(\EV^{12},\EV^{21})({\bf I}-
\Sigma{\cal V})^{-1}\Sigma
\left(\matrix{\EV^{21}\cr \EV^{12}}\right)\label{SS'}
\eea
In RHS of these equations
\bea
\Sigma= \left(\matrix{C\ES_1 C&0\cr 0& C\ES_2C}\right),
\quad\quad\quad
{\cal V} = \left(\matrix{\EV^{11}&\EV^{12}\cr \EV^{21}&\EV^{22}}\right),
\label{SigmaV}
\eea
and ${\bf I}^{\a,rs}_{\b,MN}= \delta^\a_\b\,\delta_{MN}\,\delta^{rs}$,
$r,s= 1,2$. ${\rm DET}$ is the determinant with respect to all indices.
To reach the form (\ref{SS'}) one has to use cyclicity of $\EV^{rs}$
under $r\to r+1, s\to s+1$, see above.

Let us now discuss the squeezed state solution to the equation
$|\Psi\rangle * |\Psi\rangle =|\Psi\rangle$ in the matter sector. In
order for this to be satisfied
with the above states $|S\rangle$, we must first impose
\bea
\ES_1=\ES_2=\ES'\equiv\ES\0
\eea
and then suitably normalize the resulting state.
Then (\ref{SS'}) becomes an equation for $\ES$, i.e.
\bea
\tilde \ES= \EV^{11} +(\EV^{12},\EV^{21})({\bf I}-
\Sigma{\cal V})^{-1}\Sigma
\left(\matrix{\EV^{21}\cr \EV^{12}}\right)\label{SS}
\eea
where $\Sigma,{\cal V}$ are the same as above with $\ES_1=\ES_2=\ES$.
Eq.(\ref{SS}) has an obvious (formal) solution by iteration. However
in ref. \cite{KP} it was shown that it is possible to obtain the
solution in compact form by `abelianizing' the problem.
Notwithstanding the differences with that case, it is possible
to reproduce the same trick on eq.(\ref{SS}), thanks to (\ref{commute}).
One denotes $C\EV^{rs}$ by $\EX^{rs}$ and $C\ES$ by $\ET$, and
assumes that $[\EX^{rs},\ET]=0$ (of course this has to be checked
{\it a posteriori}). Notice however that we cannot assume that
$C$ commutes with $\ES$, but we assume that $C\ES = \tilde \ES C$.
By multiplying (\ref{SS}) from the left by $C$ we get:
\bea
 \ET= \EX^{11} +(\EX^{12},\EX^{21})({\bf I}- \Sigma{\cal V})^{-1}
\left(\matrix{\ET\EX^{21}\cr \ET\EX^{12}}\right)\label{TT}
\eea
For instance $\tilde\ES \EV^{12}= \tilde\ES C C \EV^{12}= \ET \EX^{12}$,
etc. In the same way,
\bea
({\bf I}- \Sigma{\cal V})^{-1}=
\left( \matrix{{\I}-\ET\EX^{11}& -\ET\EX^{12}\cr
-\ET \EX^{21}& {\I}- \ET\EX^{11}}\right)^{-1}\0
\eea
where $\I^{\a}_{\b,MN}= \delta^\a_\b\,\delta_{MN}$. Now
all the entries are commuting matrices, so the inverse can
be calculated straight away.

From now on everything is the same as in \cite{KP,RSZ2}, therefore we
limit ourselves to a quick exposition. One arrives at an
equation only in terms of $\ET$ and $\EX
\equiv \EX^{11}$:
\bea
(\ET-\I)(\EX\ET^2- (\I+\EX)\ET +\EX)=0\label{fineq}
\eea
This gives two solutions:
\bea
&&\ET =\I\label{sol1}\\
&&\ET = \frac 1{2\EX}\left( \I +\EX - \sqrt{(\I + 3\EX)(\I-\EX)}\right)
\label{sol2}
\eea
The third solution, with a + sign in front of the square root, is not
acceptable,
as explained in \cite{RSZ2}. In both cases we see that the solution
commutes with $\EX^{rs}$. The squeezed state solution we are looking
for is, in both cases,
$\ES=C\ET$. As for (\ref{sol1}), it is easy to see that it leads to
the identity state. Therefore, from now on we will consider
(\ref{sol2}) alone.

Now, let us deal with the normalization of $|S_\perp\rangle$.
Imposing $|S_\perp\rangle\, *\,|S_\perp\rangle = |S_\perp\rangle$ we
find
\bea
{\cal N}^2 = K_2^{-1} \,{\rm DET}\, ({\bf I} - \Sigma {\cal V})^{1/2}\0
\eea
Replacing the solution in it one finds
\bea
 {\rm DET} ({\bf I} - \Sigma {\cal V}) = {\rm Det}\,
 \left( (\I -\EX)(\I + \ET)\right)
\label{det}
\eea
${\rm Det}$ denotes the determinant with respect to the indices
$\a,\b,M,N$.
Using this equation and (\ref{K2}), and borrowing from \cite{RSZ2}
the expression for $|S_\|\,\rangle$, one finally gets for the
23--dimensional tachyonic lump
\bea
|S\rangle \!&=&\! \left\{{\rm det}(1-X)^{1/2}{\rm det}
(1+T)^{1/2}\right\}^{24}
{\rm exp}\left(-\frac 12 \eta_{\bar \mu\bar \nu}\sum_{m,n\geq 1}
a_m^{\bar \mu\dagger}S_{mn}a_n^{\bar \nu\dagger}\right)|0\rangle
\otimes\label{fullsol}\\
&& \frac {A^2 (3+4a^2)}{\sqrt{2 \pi b^3}({\rm Det}G)^{1/4}}
\left( {\rm Det}(\I -\EX)^{1/2}{\rm Det}(\I + \ET)^{1/2}\right)
{\rm exp}\left(-\frac 12 \sum_{M,N\geq 0}
a_M^{\a\dagger}\ES_{\a\b,MN}a_N^{\b\dagger}\right)|\tilde 0 \rangle,\0
\eea
where $\ES= C\ET $ and $\ET$ is given by (\ref{sol2}).
The quantities in the first line are defined in ref.\cite{RSZ2} with
$\bar\mu,\bar\nu=0,
\ldots 23$ denoting the parallel directions to the lump.

The value of the action corresponding to (\ref{fullsol}) is easily
calculated
\bea
{\cal S}_\ES\!&=&\! {\EuScript K} \frac {V^{(24)}}{(2\pi)^{24}}
\left\{{\rm det}(1-X)^{3/4}{\rm det} (1+3X)^{1/4}\right\}^{24}\0\\
&&\cdot\, \frac {A^4 (3+4a^2)^2}{{2 \pi b^3}({\rm Det}G)^{1/2}} \,\,
{\rm Det} (\I -\EX)^{3/4}{\rm Det}(\I + 3\EX)^{1/4}\label{actionS}
\eea
where $V^{(24)}$ is the volume along the parallel directions and
${\EuScript K}$ is the constant of eq.(\ref{action2}).

Finally, let $\mathfrak e$ denote the energy per unit volume,
which coincides with the brane tension when $B=0$. Then one can
compute the ratio of the D23--brane energy density
${\mathfrak e}_{23}$ to the D25-brane
energy density ${\mathfrak e}_{25}$ ;
\bea
\frac {{\mathfrak e}_{23}}{ {\mathfrak e}_{25}} &\!=\!&
\frac {(2\pi)^2} {({\rm Det}G)^{1/4}}\cdot
{\EuScript{R}} \label {ratio1} \\
\ER &\!=\!&
\frac {A^4 (3+4a^2)^2}{2 \pi b^3({\rm Det}G)^{1/4}}
\frac {{\rm Det}(\I -\EX)^{3/4}{\rm Det}(\I + 3\EX)^{1/4}}
{{\rm det}(1 -X)^{3/2}{\rm det}(1 + 3X)^{1/2}}\label{ratio2}
\eea

If the quantity $\ER$ equals 1, this equation is
exactly what is expected for the ratio of a flat static D25--brane
action and a D23--brane action per unit volume in the presence
of the $B$ field (\ref{B})\cite{Oka,MRVY}. In fact the DBI Lagrangian for a flat
static Dp--brane is, \cite{SW},
\bea
{\cal L}_{DBI} = \frac 1{g_s (2\pi)^p} \sqrt { {\rm Det} (1+ 2\pi
B)}\label{DBI}
\eea
where $g_s$ is the closed string coupling.
Substituting (\ref{B}) and taking the ratio the claim follows.
By extending the methods of \cite{oku3} (see also \cite{HM1,HM2}) to the
present case, we have
indeed being able to prove in \cite{BMS2} that
\bea
\ER =1\label{R=1}
\eea
thus adding evidence to the interpretation of $|S\rangle$, given by
(\ref{fullsol}), as a D23--brane in the presence of a background $B$
field. A further confirmation of this interpretation could be obtained
from the study of the spectrum of modes leaving on the brane, which can
presumably be done along the same lines as \cite{HKw,HK,HM1,HM2,Oka}.

To end this section let us briefly discuss the generalization of the
above results to lower dimensional lumps. As remarked at the end of
section 2, every couple of transverse directions corresponding to
an eigenvalue $B_i$ of the field $B$ can be treated in the same way
as the 24--th and 25--th directions. One has simply to replace in the
above formulas $B$ with $B_i$. The derivation of the above formulas
for the case of $25-2i$ dimensional lumps is straightforward.

\section{Some effects of the $B$ field} \label{low}

In this section we would like to show that what we have
obtained so far is not merely a formal replica of the same calculation
without $B$ field, but that it significantly affects the lumps
solutions. Precisely we would like to
show that a $B$ field  has the effect of smoothing out some of the
singularities that appear in the VSFT, in particular in the low energy
limit.

In \cite{MT} it was shown that the geometry  of the
lower--dimensional lump states representing Dp-branes
is singular. This can be seen both in the zero slope limit
$\alpha' \rightarrow 0$ and as an exact result. It can be
briefly stated by saying that the midpoint of the string is
confined on the hyperplane of vanishing transverse
coordinates. It is therefore interesting to see whether
the presence of a $B$ field modifies this
situation. Moreover, as explained in the introduction, soliton solutions of
field theories defined on a noncommutative space describe Dp-branes
(\cite{GMS}, \cite{Komaba}). It is then interesting
to see if we can recover the simplest GMS soliton, using
the particular low energy limit, i.e. the  limit of \cite{SW}, that
gives a noncommutative field theory from a string theory with a $B$
field turned on.

We start with the limit of \cite{SW}, $\alpha' B \gg g$, in such a
way that $G$, $\theta$ and $B$ are kept fixed, which we represent
by means of a parameter $\epsilon$ going to $0$ as in \cite{MT}
($\alpha' \sim \epsilon^2$). We write the closed string metric
$g_{\a\b}$
as $ g\,\delta_{\a\b}$.  We could also choose to parametrize the
$\alpha' B \gg g$ condition by sending $B$ to infinity, keeping $g$
and $\alpha'$ fixed and operating a rescaling of the string modes as
in \cite{Sch}, of course at the end we get identical results.
By looking at the exponential of the 3-string field theory vertex in
the presence of a $B$ field
\bea
&&\sum_{r,s=1}^3\left(\frac 12 \sum_{m,n\geq 1}
G_{\alpha\beta}a_m^{(r)\alpha\dagger}V_{mn}^{rs}
a_n^{(s)\beta\dagger}
+ \sqrt{\alpha'}\,\sum_{n\geq 1}G_{\alpha\beta}
p_{(r)}^{\alpha}V_{0n}^{rs}
a_n^{(s)\beta\dagger} \0 \right. \\
&& \left. \quad \quad  + \,\alpha'\,\frac 12 G_{\alpha\beta}
p_{(r)}^{\alpha}V_{00}^{rs}
p_{(s)}^\beta +
\frac i2 \sum_{r<s} p_\alpha^{(r)}\theta^{\alpha\beta}
p_\beta^{(s)}\right)
\eea
we see that the limit is characterized by the rescalings
\bea
&& V_{mn}  \rightarrow  V_{mn} \0  \\
&& V_{m0}  \rightarrow  \epsilon V_{m0}  \\
&& V_{00}  \rightarrow  \epsilon^2 V_{00} \0
\eea
The dependence of $G_{\a\b}$ and $\theta^{\a\b}$ on
$g$, $\alpha'$ and $B$ is understood. We will make it explicit
at the end of our calculations in the form
\bea
G_{\a\b}  = \frac{(2\pi\alpha'B)^2}{g}\delta_{\a\b}, \quad\quad
 \theta   =  \frac{1}{B} \label{SWGB}
\eea
Substituting the leading behaviors of $V_{MN}$ in eqs.(\ref{VVmn}),
and keeping in mind that $A = V_{00} + \frac{b}{2}$, the coefficients
$\EV_{MN}^{\a\b,rs}$ become
\bea
\EV_{00}^{\a\b,rs} & \rightarrow &\, G^{\a\b}\delta^{rs}-
\frac {4}{4a^2+3}
\left(G^{\a\b} \phi^{rs} -ia \epsilon^{\a\b}\chi^{rs}\right)\\
\EV_{0n}^{\a\b,rs} & \rightarrow &\, 0\\
\EV_{mn}^{\a\b,rs} & \rightarrow &\, G^{\a\b}V_{mn}^{rs}\label{VVmnt}
\eea
We see that the squeezed state (\ref{fullsol}) factorizes in two parts:
the coefficients $\EV_{mn}^{\a\b,11}$ reconstruct the full 25
dimensional sliver, while the coefficients $\EV_{00}^{\a\b,11}$
take a very simple form
\bea
&&\ES_{00}^{\a\b}=\frac{2 |a| -1}{2 |a| +1}\, G^{\a\b} \equiv s\,
G^{\a\b} \label{expB}
\eea
In the $\epsilon\to 0$ limit we also have
\bea
&& {\rm Det}(\I -\EX)^{1/2}{\rm Det}(\I + \ET)^{1/2}\to
 \frac{4}{4a^2 + 3}\, {\rm det} (1-X)\,\,
\frac{4a}{2a+1}\, {\rm det} (1+T) \label{dets}
\eea
The complete lump state in this limit will be denoted by
$|\hat{\ES}\rangle$, and as a consequence of eq.(\ref{fullsol}) and
these equations, it will take the form
\bea
|\hat{\ES}\rangle \!&=&\! \left\{\det(1-X)^{1/2}
\det (1+T)^{1/2}\right\}^{26}
{\rm exp}\left(-\frac 12 G_{ \mu \nu}\sum_{m,n\geq 1}
a_m^{ \mu\dagger}S_{mn}a_n^{ \nu\dagger}\right)|0\rangle \otimes
\label{solB}\\
& & \frac{4a}{2a+1}\,\, \frac{b^2}{\sqrt{2\pi b^3}(\det G)^{1/4}}\,\,
{\rm exp}\left(-\frac 12 s \,
a_0^{\a\dagger}G_{\a\b}a_0^{\b\dagger}\right)| \Omega_{b, \theta}\rangle,\0
\eea
where  $\mu, \nu = 0, \dots 25$ and $G_{\mu\nu} =
\eta_{\bar\mu \bar\nu}\otimes
G_{\a\b}$. The first line of the RHS of this equation is nothing but the
sliver state $|\Xi\rangle$, which represents the D25--brane. The norm
of the lump is now regularized by the presence
of $a$ which is directly proportional to $B$: $a = -\frac{\pi^2}{A}B$.
Using
\bea
&& | x \rangle = \sqrt{\frac{2\sqrt{\det G}}{b\pi}}
\exp\left[-\frac{1}{b}x^{\a}G_{\a\b}x^{\b}
-\frac{2}{\sqrt{b}}i a_0^{\a\dagger} G_{\a\b}x^{\b}
+\frac{1}{2}a_0^{\a\dagger}G_{\a\b}a_0^{\b\dagger}
\right]
|\Omega_{b, \theta} \rangle
\eea
we can calculate the projection onto the basis of position
eigenstates of the transverse part of the lump state
\bea
\langle x | e^{-\frac{s}{2}(a_0^{\dagger})^2}|\Omega_{b, \theta} \rangle & = &
\sqrt{\frac{2\sqrt{\det G}}{b\pi}} \frac{1}{1+s}\,
 e^{-\frac{1-s}{1+s}\frac{1}{b}x^{\a}x^{\b}G_{\a\b}} \0 \\
& = & \sqrt{\frac{2\sqrt{\det G}}{b\pi}} \frac{1}{1+s}\,
      e^{-\frac{1}{2|a|b}x^{\a}x^{\b}G_{\a\b}} \label{xproj}
\eea
The transverse part of the lump state in the $x$ representation is then
\bea
\langle x |\hat{\ES}_\perp \rangle
= \frac{1}{\pi}\, e^{-\frac{1}{2|a|b}x^{\a}x^{\b}G_{\a\b}} |\Xi_\perp
\rangle\label{regul}
\eea

Using now the form (\ref{SWGB}) of $G_{\a\b}$ and $\theta^{\a\b}$
and the explicit expression for $a$ in terms of $g$
and $\alpha'$, \cite{BMS1}
\bea
a = \frac{\theta}{4A}\sqrt{\det G} =
- \frac{2 \pi^2 (\alpha')^2 B}{b\, g}
\eea
we obtain the simplest soliton solution
of \cite{GMS} (see also \cite{Komaba} and references therein):
\bea
 e^{-\frac{1}{2|a|b}x^{\a}x^{\b}G_{\a\b}} &\rightarrow &
e^{-\frac{x^{\a}x^{\b}\delta_{\a\b}}{|\theta|}}
\eea
which corresponds to the $|0\rangle\langle 0|$ projector in
the harmonic oscillator Hilbert space (see Introduction), and
is a projector on a space endowed with a Moyal product.

In this way the $B$ field provides a regularization of (\ref{regul}),
as compared to \cite{MT}. This beneficial effect of the $B$ field
is confirmed by the fact that the projector (\ref{solB}) is no longer
annihilated by $x_0$
\bea
x_{0}\,\, {\rm exp}\left(-\frac 12
s a_0^{\a\dagger}G_{\a\b}a_0^{\b\dagger}\right)
| \Omega_{b, \theta}\rangle
& = & i\,\frac{\sqrt{b}}{2}\,(a_0 - a_0^{\dagger})\,
{\rm exp}\left(-\frac 12
s a_0^{\a\dagger}G_{\a\b}a_0^{\b\dagger}\right)
| \Omega_{b, \theta}\rangle \0 \\
& =& -
i\,\frac{\sqrt{b}}{2}\,
\left[\frac{4a}{2a+1}\right]a_0^{\dagger}\, {\rm exp}\left(-\frac 12
s a_0^{\a\dagger}G_{\a\b}a_0^{\b\dagger}\right)
| \Omega_{b, \theta}\rangle \0
\eea
Therefore, in the low energy limit, the singular structure
found in \cite{MT} has disappeared in the presence of a nonvanishing
$B$ field. This is actually not true only
in the low energy limit, but is an exact result, as was shown
in \cite{BMS2}.

\section{More lumps in VSFT}

In the two previous sections we have constructed a 23--dimensional
lump solution, which we have interpreted as a D23--brane. In the low
energy limit this solution in the coordinate basis, turned out to be
the simplest (two--dimensional) GMS soliton multiplied by a translational
invariant solution which represents the D25--brane.
The question we want to deal with here is whether there are other
lump solutions that correspond to the higher order GMS solitons.
The answer is affirmative. We will construct an infinite sequence
of them, denoted $|\Lambda_n\rangle$. These new star algebra projectors
are D23--branes, constructed out of
(\ref{fullsol}) and parallel to it. In the
low energy limit they give rise to the full series of GMS solitons.
We will construct them and prove that they satisfy the remarkable
identities (\ref{nstarm},\ref{nm}).

In order to construct these new solutions we need a new ingredient, given
by the Fock space projectors similar to those introduced in
\cite{RSZ3}. We define them only along the transverse directions
\bea
\rho_1 \!&=&\! \frac 1{(\I +\ET)(\I-\EX)} \left[ \EX^{12} (\I-\ET\EX)
+\ET (\EX^{21})^2\right]\label{rho1}\\
\rho_2 \!&=&\! \frac 1{(\I +\ET)(\I-\EX)} \left[ \EX^{21} (\I-\ET\EX)
+\ET (\EX^{12})^2\right]\label{rho2}
\eea
They satisfy
\bea
\rho_1^2 = \rho_1,\quad\quad \rho_2^2 = \rho_2, \quad\quad
\rho_1+\rho_2 = \I\label{proj12}
\eea
i.e. they project onto orthogonal subspaces. Moreover, if we
use the superscript $^T$ to denote transposition with respect to
the indices $N,M$ and $\a,\b$, we have
\bea
\rho_1^T=\tilde\rho_1 = C\rho_2 C,\quad\quad
 \rho_2^T=\tilde\rho_2 = C\rho_1 C.\label{rhorels}
\eea
and
\bea
&&\rho_i^\dagger = \rho_i, \quad {\rm i.e.}\quad \rho_i^*= \tilde\rho_i,
\quad i=1,2\0\\
&&\tau \rho_i = \tilde \rho_i \tau, \quad \quad i=1,2\0
\eea
where $^*$ denote complex conjugation and $^\dagger= ^{*T}$. Moreover
$\tau$ is the matrix $\tau = \{ \tau_\a{}^\b\}=$
\mbox{ \small $\left(\begin{array}{cc}1 & 0 \\ 0
& -1 \end{array}\right)$ }. We recall that in the absence of the $B$ field,
it has been shown that $\rho_1,\rho_2$ projects out half the string modes,
\cite{RSZ3,Moeller}.

With all these ingredients we can now move on, give a precise
definition of the $|\Lambda_n\rangle$ states and demonstrate the
properties announced above.

To define the states $|\Lambda_n\rangle$ we start from the lump
solution (\ref{fullsol}). I.e. we take $|\Lambda_0\rangle= |\ES\rangle$.
However, in the following, we will limit ourselves only to the transverse
part of it, the parallel one being universal and
irrelevant for our construction. We will denote the transverse part
by $|\ES_\perp\rangle$.

First we introduce two `vectors' $\xi=\{\xi_{N\a}\}$ and $\zeta =
\{\zeta_{N\a}\}$, which are chosen to satisfy the conditions
\bea
\rho_1 \xi =0,\quad\quad \rho_2 \xi =\xi, \quad\quad
{\rm and}\quad \rho_1 \zeta =0,\quad\quad \rho_2\zeta=\zeta,
\label{xizeta}
\eea

Next we define
\bea
\x = (a^\dagger \tau \xi)\, (a^\dagger C \zeta)=
(a_N^{\a\dagger} \tau_\a{}^\beta \xi_{N\b})
(a_N^{\a\dagger}C_{NM}\zeta_{M\a})   \label{xbf}
\eea
and introduce the Laguerre polynomials $L_n(z)$, of the generic
variable $z$. The definition of $|\Lambda_n\rangle$ is as follows
\bea
|\Lambda_n\rangle = (-\kappa)^n L_n\Big(\frac{\x}{\kappa}\Big)
|\ES_\perp\rangle\label{Lambdan}
\eea
As part of the definition of $|\Lambda_n\rangle$ we require the two
following conditions to be satisfied
\bea
\xi^T \tau\frac 1{\I-\ET^2}\zeta = -1 ,\quad\quad
\xi^T \tau\frac {\ET}{\I-\ET^2}\zeta= -\kappa
\label{cond}
\eea

Hermiticity for $|\Lambda_n\rangle$ requires that
\bea
(a \tau \xi^*)(aC \zeta^*) = (a\tau C\xi)(a \zeta)\label{hermit}
\eea
This condition admits the solution
\bea
\zeta = \tau \xi^*\label{real}
\eea
which we will assume throughout the rest of the paper, even though
it will be left implicit for notational simplicity. Eq.(\ref{real}) is
compatible with the conditions (\ref{xizeta}) and (\ref{cond}), see \cite{BMS3}.
As a consequence of (\ref{real}), the LHS's of both equations (\ref{cond})
are real, so $\kappa$ must be real too. Let us show this for instance for the first
equation, since for the second no significant modification is needed:
\bea
\left(\xi^T \tau\frac 1{\I-\ET^2}\zeta\right)^*=
\xi^{T*} \tau\frac 1{\I-(\ET^*)^2}\zeta^* =
\zeta^T \frac 1{\I-(\ET^*)^2}\tau\xi =
\xi^T \tau\frac 1{\I-(\ET^\dagger)^2}\zeta=
\xi^T \tau\frac 1{\I-\ET^2}\zeta \0
\eea
where the second equality is obtained by replacement of (\ref{real}),
and the third by transposition.

The proof that (\ref{nstarm},\ref{nm}) are satisfied was given in \cite{BMS3}.

Before we pass to the low energy limit, let us make a comment on the
definition of $|\Lambda_n\rangle$, wherein a central role is played by
the Laguerre polynomials. While the true rationale of this role eludes
us, it is possible to prove that the form of the definition
$|\Lambda_n\rangle$ (together with (\ref{xizeta},\ref{cond})
is not only sufficient for (\ref{nstarm},\ref{nm}) to be true,
but also necessary.
The case $|\Lambda_1\rangle = (\x - \kappa)|\ES_\perp\rangle $
was discussed in \cite{RSZ3}. The next most complicated state is
\bea
(\a + \b \x + \gamma \x^2) |\ES_\perp\rangle \label{try}
\eea
The conditions this state has to satisfy in order to define a
$|\Lambda_{2}\rangle$ that obeys (\ref{nstarm},\ref{nm})
for $n=0,1,2$ are, given (\ref{xizeta},\ref{cond}), by the following
relations
\bea
-2(\a)^{1/2} = \b, \quad\quad  \gamma = \frac12
\eea
Then, putting $\a = \kappa$
\bea
|\EP'\rangle =\Big(\kappa^2 - 2 \kappa\x + \frac 12 \x^2\Big)
|\ES_\perp\rangle\label{try'}
\eea
The polynomial in the RHS is nothing but the second Laguerre
polynomial of $\x/\kappa$ multiplied by $\kappa^2$. In fact using
Mathematica it is easy to extend this analysis for $n$ as large as one
wishes.

Finally let us remark that the relations demonstrated in this section,
in particular (\ref{nstarm},\ref{nm}), are true for any value of $B$,
therefore also for $B=0$.

\section{The GMS solitons}

In order to analyze the same limit as in section (\ref{low}) for a generic
$|\Lambda_n\rangle$,
first of all we have to find the low energy limit of the
projectors $\rho_1, \rho_2$. In this limit these two projectors
factorize
into the zero mode and non--zero mode part. The former is given by
\bea
(\rho_1)_{00}^{\a\b}\rightarrow \frac 12 \Big[G^{\a\b}+
i \epsilon^{\a\b}\Big],
\quad\quad
(\rho_2)_{00}^{\a\b}\rightarrow \frac 12 \Big[G^{\a\b}-
i \epsilon^{\a\b}\Big],\label{rholimit}
\eea

Now, we take, in the definition (\ref{xbf}),
$\xi= \hat \xi+ \bar\xi$ and
$\zeta= \hat \zeta + \bar\zeta$, where
$\bar\xi, \bar\zeta$ are such that they vanish in the
limit $\a' \to 0$. Then we make the choice
$\hat\xi_n = \hat \zeta_n = 0, \,\,\, \forall n>0$ and determine
$\hat \xi$ and $\hat \zeta$ in such a way that
eqs.(\ref{xizeta}, \ref{hermit}) and (\ref{cond}) are satisfied in the
limit $\a'\to 0$. We are assuming here that there exist solutions
of the problem at $\alpha'\neq 0$ that take precisely this specific form
when $\alpha'\to 0$. This is a plausible assumption since the
$\alpha'$ dependence is smooth in all the involved quantities.
In any case it is not hard to construct examples of this fact: for instance
$\xi =\rho_2\hat \xi$ satisfies the above requirements to zeroth and
first order of approximation in $\epsilon$. More complete examples are provided in \cite{BMS3}.

Now, in the field theory limit the conditions
(\ref{xizeta}) become
\bea
\hat\xi_{0,24} + i \hat\xi_{0,25}=0, \quad\quad \hat\zeta_{0,24}
+ i \hat\zeta_{0,25}=0,
\label{0cond}
\eea
From now on we set $\hat\xi_0 = \hat\xi_{0,25}= - i \hat\xi_{0,24}$
and, similarly,
$\hat\zeta_0 = \hat\zeta_{0,25}= - i\hat \zeta_{0,24}$.
The conditions (\ref{cond}) become
\bea
&&\xi^T\tau \frac 1{\I-\ET^2}\zeta \rightarrow - \frac{1}{1-s^2}
\frac 2{\sqrt{\det G}}{\hat\xi_0\hat\zeta_0} =-1\label{cond1}\\
&&\xi^T\tau \frac {\ET}{\I-\ET^2}\zeta \rightarrow -\frac{s}{1-s^2}
\frac 2{\sqrt{\det G}}{\hat\xi_0\hat\zeta_0}  =-\kappa\label{cond2}
\eea
Compatibility requires
\bea
\frac{2\hat\xi_0\hat\zeta_0}{\sqrt{\det G}}= 1-s^2,\quad\quad \kappa= s
\label{comp}
\eea
At the same time
\bea
(\xi \tau a^\dagger)(\zeta C a^\dagger) \rightarrow -
{\hat\xi_0\hat\zeta_0}
((a_0^{24\dagger})^2+(a_0^{25\dagger})^2)= -\frac {\hat\xi_0\hat\zeta_0}
{\sqrt{\det G}} a_0^{\a\dagger} G_{\a\b}a_0^{\b\dagger}\label{xitaua}
\eea
Hermiticity requires that the product $\hat\xi_0
\hat\zeta_0= |\hat\xi_0|^2$, in accordance with (\ref{cond1},\ref{cond2}).
The solutions found in this way can be referred to
as the {\it factorized solutions}, since, as will become clear
in a moment, they realize the factorization of the star product
into the Moyal $\star$ product and Witten's $\ast$ product.
In order to be able to compute $\langle x| \Lambda_n\rangle$ in the
field theory limit, we have to evaluate first
\bea
\langle x | \left(a_0^{\a\dagger} G_{\a\b}a_0^{\b\dagger}\right)^k\,
e^{-\frac{s}{2} a_0^{\a\dagger} G_{\a\b}a_0^{\b\dagger} }
|\Omega_{b, \theta} \rangle &=&
(-2)^k \frac {d^k}{d s^k} \left( \langle x |
e^{-\frac{s}{2} a_0^{\a\dagger}
G_{\a\b}a_0^{\b\dagger} }|\Omega_{b, \theta} \rangle\right)
\label{xkproj}\\
&=& (-2)^k \frac {d ^k}{d s^k} \left( \sqrt{\frac{2\sqrt{\det G}}{b\pi}}
\frac{1}{1+s}\,
 e^{-\frac{1-s}{1+s}\frac{1}{b}x^{\a}G_{\a\b}x^{\b}}\right) \0
\eea
An explicit calculation gives
\bea
&& \frac {d^k}{d s^k} \left(\frac{1}{1+s}\,
 e^{-\frac{1-s}{1+s}\frac{1}{b}x^{\a}x^{\b}G_{\a\b}}\right) =\label{dsk}\\
&&\quad\quad= \sum_{l=0}^k \sum_{j=0}^{k-l}
\frac {(-1)^{k+j}}{(1-s)^j(1+s)^{k+1}}
\frac{k!}{j!} \left(\matrix{k-l-1\cr j-1}\right)\, \langle x,x \rangle^j
e^{-\frac{1}{2}\langle x,x\rangle}\0
\eea
where it must be understood that, by definition, the binomial
coefficient \mbox{ \small $\left(\matrix{-1\cr -1}\right)$} equals 1.
Moreover we have set
\bea
\langle x,x \rangle = \frac{1}{ab}x^{\a} G_{\a\b}x^{\b} =
\frac{ 2 r^2 }{\theta}
\eea
with $r^2 = x^\a x^\b \delta_{\a\b}$.

Now, inserting (\ref{dsk}) in the definition of $|\Lambda_n\rangle$,
we obtain after suitably reshuffling the indices:
\bea
 \langle x| (-\kappa)^n L_n\Big(\frac{\x}{\kappa}\Big)
e^{-\frac 12 s a^{\a \dagger}_0 G_{\a\b} a^{\b \dagger}_0}
|\Omega_{b, \theta}\rangle
 & \to & \langle x | (-s)^n L_n \Big(-\frac{1-s^2}{2s}\,\,
a_{0}^{\a\dagger}G_{\a\b}a_{0}^{\b\dagger}\Big)
e^{-\frac 12 s a_0^{\a \dagger} G_{\a\b} a_0^{\b \dagger}}|\Omega_{b, \theta}\rangle \0\\
&=&
\frac{(-s)^n}{(1+s)} \sum_{j=0}^n\sum_{k=j}^n \sum_{l=j}^k
\left(\matrix{n\cr k}\right)\left(\matrix{l-1\cr j-1}\right)\frac 1{j!}
\frac {(1-s)^k}{(1+s)^{j}s^k}\0\\
&\cdot& (-1)^j \langle x,x \rangle^j\,
e^{-\frac{1}{2}\langle x,x \rangle}
\sqrt{\frac{2\sqrt{\det G}}{b\pi}} \label{xlambda1}
\eea
The expression can be evaluated as follows. First one uses the result
\bea
\sum_{l=j}^k \left(\matrix{l-1\cr j-1}\right) =
\left(\matrix{k\cr j}\right) \label{c1}
\eea
Inserting this into (\ref{xlambda1}) one is left with the following
summation, which contains an evident binomial expansion,
\bea
\sum_{k=j}^n   \left(\matrix{n\cr k}\right)\left(\matrix{k\cr j}\right)
\left(\frac{1-s}{s}\right)^k=
\left(\matrix{n\cr j }\right) \frac{(1-s)^j}{s^n}\label{c3}
\eea
Replacing this result into (\ref{xlambda1}) we obtain
\bea
\langle x| (-\kappa)^n L_n\left(\frac{\x}{\kappa}\right)
e^{-\frac 12 s a^{\a \dagger}_0 G_{\a\b} a^{\b \dagger}_0}
|\Omega_{b, \theta}\rangle &\rightarrow&
\frac{2|a|+1}{4|a|}\sqrt{\frac {2\sqrt{\det G}}{b \pi }}(-1)^n
\sum_{j=0}^n
\left(\matrix{n\cr j}\right)\frac 1{j!} \,
\left(- \frac{2r^2}{\theta}\right)^j
e^{-\frac {r^2}{\theta}}\0
\eea
Recalling now
that the definition of $|\hat \ES\rangle$ includes an additional
numerical factor (see eq.(\ref{solB})), we finally obtain
\bea
\langle x| \Lambda_n\rangle \rightarrow \langle x|
 \hat\Lambda_n\rangle&=& \frac 1{\pi}(-1)^n
\sum_{j=0}^n
\left(\matrix{n\cr j}\right)\frac 1{j!} \,
\left(- \frac{2r^2}{\theta}\right)^j
e^{-\frac {r^2}{\theta}} |\Xi\rangle \0\\
&=& \frac 1{\pi}(-1)^n\,L_n\left(\frac{2r^2}{\theta}\right)
\,e^{-\frac {r^2}{\theta}}|\Xi\rangle
\eea
as announced in section 6. The coefficient
in front of the sliver $|\Xi\rangle$ is the $n-th$ GMS
solution. Strictly speaking there is a discrepancy between
these coefficients and the corresponding GMS soliton, given
by the normalizations which differ by a factor of $2\pi$.
This can be traced back to the traditional normalizations
used for the eigenstates $|x\rangle$ and $|p\rangle$ in the
SFT theory context and in the Moyal context, respectively.
This discrepancy can be easily dealt with a simple redefinition.

\section{VSFT star product and Moyal product}

In the previous section we have shown that the low energy limit of
$\langle x| \hat\Lambda_n\rangle$ factorizes into the product of the sliver state
and $\psi_n(x,y)$, see (\ref{solitons}). This means, on the one hand,
that the GMS
solitons are the low energy remnants of corresponding D--branes in VSFT,
and, on the other hand, that, for this type of
solutions, the VSFT star product factorizes into Witten's
star product and
the Moyal $\star$ product. But, actually, much more can be said about
the correspondence between the states $| \hat\Lambda_n\rangle$ and the solitons
of noncommutative field theories with polynomial interaction.

We recall from section 6 that the latter are very elegantly
constructed in terms of harmonic oscillators eigenstates
$|n\rangle$. In particular the $\psi_n(x,y)$ solutions correspond
to projectors $P_n=|n\rangle\langle n|$, via the Weyl transform.
The correspondence is such that the operator product in the Hilbert space
corresponds to the Moyal product in $(x,y)$ space.
Therefore we can formalize the following correspondence
\bea
\matrix{|\Lambda_n\rangle &\longleftrightarrow&
P_n
&\longleftrightarrow&\psi_n (x,y)  \cr
|\Lambda_n\rangle * |\Lambda_{n'}\rangle &\longleftrightarrow&
P_n P_{n'}&\longleftrightarrow &\psi_n \star \psi_{n'}}\label{c2}
\eea
where $\star$ denotes the Moyal product. Moreover
\bea
\langle \Lambda_n|\Lambda_{n'}\rangle &\longleftrightarrow&
{\rm Tr} (P_nP_{n'}) \,\longleftrightarrow\,
\int dxdy \,\psi_n(x,y)\psi_{n'}(x,y)
\eea
up to normalization (see (\ref{nm})).
This correspondence seems to indicate that the Laguerre polynomials
hide a universal structure of these noncommutative algebras.

This parallelism can actually be pushed still further. In fact we can
easily construct the correspondents of the operators $|n\rangle\langle m|$.
Let us first define
\bea
X= a^\dagger \tau \xi \, \quad\quad Y = a^\dagger C\zeta\label{XY}
\eea
so that $\x = XY$. The definitions we are looking for are as follows
\bea
|\Lambda_{n,m}\rangle \,&=&\, \sqrt \frac {n!}{m!} (-\kappa)^n \,
Y^{m-n} L_n^{m-n}
\left(\frac \x \kappa\right)|\ES_\perp\rangle,
\quad\quad n\leq m \label{nm1}\\
|\Lambda_{n,m}\rangle \,&=&\, \sqrt \frac {m!}{n!} (-\kappa)^m \,
X^{n-m} L_m^{n-m}
\left(\frac \x \kappa\right)|\ES_\perp\rangle,
\quad\quad n\geq m\label{nm2}
\eea
where $L_n^{m-n}(z) = \sum_{k=0}^m
\left(\matrix {m\cr n-k}\right) (-z)^k/k!$.
With the same techniques as in the previous sections one can prove that
\bea
|\Lambda_{n,m}\rangle * |\Lambda_{r,s}\rangle = \delta_{m,r}|\Lambda_{n,s}\rangle
\label{LL}
\eea
for all natural numbers $n,m,r,s$. It is clear that the previous states
$|\Lambda_n\rangle$ coincide with $|\Lambda_{n,n}\rangle$. In view of (\ref{LL}),
we can extend the correspondence (\ref{c2}) to $|n\rangle\langle m| \leftrightarrow
|\Lambda_{n,m}\rangle $. Therefore, following \cite{GMS}, \cite{Komaba},
we can apply to the construction of projectors in the VSFT star algebra the
solution generating technique, in the same way as in the harmonic oscillator
Hilbert space ${\cal H}$. Naturally in this case we do not have any guarantee
that all the projectors are recovered in this way.

\appendix
\section{Appendix}

This appendix is devoted to a direct analytic proof of eqs.(\ref{UU})
and (\ref{UU'}). Let us start from the latter.

{\bf Proof of eq.(\ref{UU'})}. It is convenient to rewrite it as follows
\beq
\sum_{k=0}^\infty\tilde U_{nk} \,\tilde U_{km} = \delta_{n0}\delta_{m0}
+ \sum_{k=0}^\infty\tilde U_{nk}^{(2)} \,\tilde U_{km}^{(1)}\label{UU1}
\eeq
since, in the range $0\leq n,m<\infty$, we have $\tilde U_{km}^{(1)}=
\delta_{n0}\delta_{m0} + \tilde U_{km}^{(2)}$ and $\tilde U_{0m}^{(1)}=
\delta_{m0}$. Therefore we have to compute
\bea
\sum_{k=0}^\infty\tilde U_{nk}^{(2)} \,\tilde U_{km}^{(1)} &=&
 \oint\frac {dz}{2\pi i} \frac 1{z^{n+1}}
\oint\frac {d\zeta}{2\pi i}\oint\frac {d\theta}{2\pi i}
\oint\frac {dw}{2\pi i} \frac 1{w^{m+1}} \sum_{k=0}^{\infty}
\,\frac 1{(\zeta\theta)^{k+1}} \,\frac{f(z)}{f(\zeta)} \frac {f(\theta)}{f(w)}
\cdot\0\\
&\cdot& \left(\frac 1{1+z\zeta } -\frac {\zeta}{\zeta-z}\right)
\left(\frac 1{1+\theta w} -\frac w{w-\theta}\right)\label{UU2}
\eea
Here we have already exchanged the summation over $k$ with integrals, which
is allowed only under definite convergence conditions. The latter are
guaranteed if $|\zeta\theta|>1$, in which case
\beq
\sum_{k=0}^{\infty} \frac 1{(\zeta\theta)^{k+1}} = \frac 1{\theta\zeta-1}
\label{ksum}
\eeq
Now, we recall that, from the definition of $\tilde U^{(1)},\tilde U^{(2)}$,
we have $|z|<|\zeta|, |\theta|>|w|$. In order to comply with the condition
$|\zeta\theta|>1$ we choose to deform the $\theta$ contour while keeping
the $\zeta$ contour fixed. In doing so we have to be careful to avoid
possible singularities in $\theta$. These are poles at $\theta = w,
-\frac 1w $ and branch cuts at $\theta = \pm i$, due to the
$f(\theta)$ factor. One can deform the $\theta$ contour in such a way as
to keep the pole at $-\frac 1{w}$ external to the contour, since the $w$
contour is as small as we wish around the origin. But, of course, one
cannot avoid the branch points at $\theta = \pm i$. To make sense of
the operation we introduce a regulator $K>1$ and modify the integrand
by modifying $f(\theta)$
\beq
f(\theta) \to f_K(\theta) =
\left(\frac {K +i \theta}{K- i\theta}\right)^{\frac 23}\0
\eeq
We will take $K$ as large as needed and eventually move back to $K=1$.
Under these conditions we can safely perform the summation over $k$ in
(\ref{UU2}) and make the replacement (\ref{ksum}) in the integral.

As the next step we carry out the $\theta$ integration, which reduces to the
contribution from the simple poles at $\theta = w$ and $\theta =
\frac 1\zeta$. The RHS of (\ref{UU2}) becomes
\bea
&=& \oint\frac {dz}{2\pi i} \frac 1{z^{n+1}}
\oint\frac {d\zeta}{2\pi i}
\oint\frac {dw}{2\pi i} \frac 1{w^{m+1}} \left[
\frac{f(z)}{f(\zeta)} \frac {f_K(1/\zeta)}{f(w)}
\left(\frac 1{1+z\zeta } -\frac {\zeta}{\zeta-z}\right)
\left(\frac 1{w+\zeta } -\frac {w}{\zeta w-1}\right)\right.\0\\
&&~~~~~~~~~~~ +\left.\frac{f(z)}{f(\zeta)}\frac {w}{\zeta w-1}
\left(\frac 1{1+z\zeta } -\frac {\zeta}{\zeta-z}\right) \right]\label{UU3}
\eea
The first line corresponds to the contribution from the pole at $\theta=
\frac 1\zeta$, while the second comes from the pole at $\theta =w$.

Next we wish to integrate with respect to $\zeta$. The singularities trapped
within the $\zeta$ contour of integration are the poles at $\zeta =z,-w$
(not the poles at $\zeta= \frac 1w, -\frac 1z$). Since above we had
$K>|\theta|>\frac 1{|\zeta|}$, it follows that $|\zeta|>\frac 1K$. Therefore
also the branch points at $\zeta = \pm \frac iK$ of $f_K(1/\zeta)$ are trapped
inside the $\zeta$ contour and we have to compute the relevant contribution
to the integral. In the integrand of (\ref{UU3}) we have two cuts in
$\zeta$. One is the cut we have just mentioned, let us call it ${\mathfrak c}_{1/K}$
and let us fix it to be the semicircle of radius $1/K$ at the RHS of
the imaginary axis; the contour that surrounds it excluding all the other
singularities will be denoted $C_{1/K}$. The other cut, due to $f(\zeta)$,
with branch points at $\zeta =\pm i$, will be denoted ${\mathfrak c}_1$; the contour
that surrounds it excluding all the other
singularities will be denoted $C_1$.

After these lengthy preliminaries let us carry out the integration over
$\zeta$. We get
\bea
&=& \oint\frac {dz}{2\pi i} \frac 1{z^{n+1}}
\oint\frac {dw}{2\pi i} \frac 1{w^{m+1}} \left[ \frac{f(1/z)}{f(w)}
\left(\frac{zw}{zw-1} - \frac z{z+w}\right) + \frac {f(z)}{f(1/w)}
\left(\frac 1{1-zw } -\frac {w}{w+z}\right)\right.\0\\
&&~~~~~~~~~~~~~\left.
+ \oint_{C_{1/K}}\frac{d\zeta}{2\pi i}  (\ldots)
+ \frac {zw}{1-zw}\right]\label{UU4}
\eea
The first two terms in square brackets come from the contribution of
the poles at $\zeta= z$ and $\zeta = -w$ from the first line in (\ref{UU3}),
respectively. The symbol $(\ldots)$ represents the integrand contained
within the square brackets in the first line of (\ref{UU3}). Finally
the last term in (\ref{UU4}) is the contribution coming from the second
line of (\ref{UU3}) due to the pole at $\zeta =z$. We notice
that
\beq
\frac{zw}{zw-1} - \frac z{z+w} = \frac w{z+w}- \frac{1}{1-zw}\0
\eeq
but of course the problem here is how to evaluate the integral around
the cut. Fortunately this can be reduced to an evaluation of contributions
from poles. To see this, we first recall the properties of $f(z)$.
It is easy to see that
\beq
f(1/z) = f(-z) \quad{\rm and}\quad f(-z)= 1/f(z)\label{fprop}
\eeq
Therefore, in the limit $K\to 1$, the factor $f_K(1/\zeta)/f(\zeta)$
tends to $(f(-\zeta))^2$. As a consequence, in the same limit, the integral
of $(\ldots)$ around the cut ${\mathfrak c}_{1/K}$ is the same as the integral around
the cut ${\mathfrak c}_1$, and each equals one--half the integral around both contours,
in other words each equals one--half the integral about a contour that
surrounds both cuts and exclude all the other singularities (which
are poles). By a well-known argument, the latter integral equals the
negative of the integral of $(\ldots)$ about all the remaining singularities
in the complex $\zeta$--plane. This is easy to compute. The
remaining singularities are poles around $\zeta = z,-w,-1/z,1/w$. Notice
that there is no singularity at $\zeta =\infty$.
Carrying out this calculation explicitly we get
\bea
&=& \oint\frac {dz}{2\pi i} \frac 1{z^{n+1}}
\oint\frac {dw}{2\pi i} \frac 1{w^{m+1}}\left\{
\frac{f(1/z)}{f(w)} \left(\frac w{z+w}- \frac{1}{1-zw}\right)
+\frac {f(z)}{f(1/w)}  \left(\frac 1{1-zw } -\frac {w}{w+z}\right)\right.\0\\
&& -\frac 12 \left[ \frac{f(1/z)}{f(w)} \left(\frac w{z+w}- \frac{1}{1-zw}\right)
+\frac {f(z)}{f(1/w)}  \left(\frac 1{1-zw } -\frac {w}{w+z}\right)\right.
\label{UU5}\\
&& \left.\left.+\frac{f(1/z)}{f(w)} \left(\frac w{z+w}- \frac{1}{1-zw}\right)
+\frac {f(z)}{f(1/w)}  \left(\frac 1{1-zw } -\frac {w}{w+z}\right)\right]
+ \frac {zw}{1-zw}\right\}\0
\eea
The terms in square brackets represent the contribution from the cut
${\mathfrak c}_{1/K}$ and come from the simple poles at
$ \zeta = z, -w, -1/z, 1/w$, respectively. All the terms cancel out except the last in the third line.
So the RHS of (\ref{UU2}) reduces to
\beq
= \oint\frac {dz}{2\pi i} \frac 1{z^{n+1}}
\oint\frac {dw}{2\pi i} \frac 1{w^{m+1}} \sum _{k=1}^\infty (zw)^k =
\delta_{nm}, \quad\quad n,m\geq 1\label{UU6}
\eeq
This complete the proof of (\ref{UU}). We remark that we could have integrated first
with respect to $\zeta$ and then with respect to $\theta$. The procedure
is somewhat different, but the final result is the same. We also point out
that there may be other equivalent ways to derive (\ref{UU}).

{\bf Proof of eq.(\ref{UU})}. It is convenient to rewrite $U_{nm}$ in an
alternative form compared to (\ref{Anm}). We start by replacing
in eq.(\ref{neumann})
\beq
f'_a(z)
\frac{1}{(f_a(z)-f_b(w))^2}f'_b(w) = -\d_z \frac{1}{f_a(z)-f_b(w)}
f_b'(w)\0
\eeq
and integrating by part. We decompose the resulting expression as in
eq.(\ref{decomp}). After some algebra one gets
\beq
U_{nm} = \sqrt {\frac nm} \oint\frac {dz}{2\pi i} \frac 1{z^{n+1}}
\oint\frac {dw}{2\pi i} \frac 1{w^{m+1}} \frac {g(z)}{g(w)}
\left( \frac 1{1 +zw} - \frac w{w-z}\right)\label{newU}
\eeq
where
\beq
g(z) = \frac 1z (1+iz)^{\frac 23} (1-iz)^{\frac 43}\label{g}
\eeq
This function satisfies
\beq
g(1/z) = g(-z) \label{gprop}
\eeq
which corresponds to the first of eqs.(\ref{fprop}). There is no analog
of the second.

In order to prove eq.(\ref{UU}) we have to evaluate
\bea
\sqrt {\frac mn}\,\sum_{k=1}^\infty U_{nk} U_{km} &=&
 \oint\frac {dz}{2\pi i} \frac 1{z^{n+1}}
\oint\frac {d\zeta}{2\pi i}\oint\frac {d\theta}{2\pi i}
\oint\frac {dw}{2\pi i} \frac 1{w^{m+1}} \sum_{k=1}^{\infty}
\,\frac 1{(\zeta\theta)^{k+1}} \,\frac{g(z)}{g(\zeta)}
\frac {g(\theta)}{g(w)} \cdot\0\\
&\cdot& \left(\frac 1{1+z\zeta } -\frac {\zeta}{\zeta-z}\right)
\left(\frac 1{1+\theta w} -\frac w{w-\theta}\right)\label{UU7}
\eea
The structure is the same as in (\ref{UU2}), except for the substitution
$f \to g$ and for the fact that now the summation over $k$ starts from 1.
We will thus proceed as above while paying attention to the differences.
Using
\beq
\sum_{k=1}^{\infty} \frac 1{(\zeta\theta)^{k+1}} = \frac 1{\zeta\theta}
\frac 1{\theta\zeta-1} \label{ksum1}
\eeq
instead of (\ref{ksum}), we see that, when integrating over $\theta$
we have to take into account the pole at $\theta = 0$. The result is
\bea
&=& \oint\frac {dz}{2\pi i} \frac 1{z^{n+1}}
\oint\frac {d\zeta}{2\pi i}
\oint\frac {dw}{2\pi i} \frac 1{w^{m+1}} \left[
\frac{g(z)}{g(\zeta)} \frac {g_K(1/\zeta)}{g(w)}
\left(\frac 1{1+z\zeta } -\frac {\zeta}{\zeta-z}\right)
\left(\frac 1{w+\zeta } -\frac {w}{\zeta w-1}\right)\right.\0\\
&&~~+\left.\frac{f(z)}{f(\zeta)}\frac {w}{\zeta w-1}
\left(\frac 1{1+z\zeta } -\frac {\zeta}{\zeta-z}\right)
+ \frac {g(z)}{\zeta g(\zeta) g(w)} \left(\frac 1{1+z\zeta }
-\frac {\zeta}{\zeta-z}\right)\frac {1+w^2}{w}\right]\label{UU8}
\eea
The last contribution comes precisely from the double pole at $\theta=0$.

Next let us untegrate over $\zeta$. There is no singularity at $\zeta=0$
or $\zeta =\infty$, as one may have suspected. Let us deal first with
the first line in eq.(\ref{UU8}). This is exactly the first line
of (\ref{UU3}), except for the substitution $f \to g$. We proceed in the
same way as above, but with some additional care because we cannot use
the analog of the second eq.(\ref{fprop}). However we remark that
\beq
\frac {g_K(1/\zeta)}{g(\zeta)} = \frac {f_K(1/\zeta)}{f(\zeta)}
\frac {(\zeta K -i)^2}{(1 - i \zeta)^2}\label{UU9}
\eeq
Now we have recovered the same structure as in (\ref{UU3}) except
for the last factor in the RHS of (\ref{UU9}), i.e. at the price of
bringing into the game a double pole at $\zeta= -i$. Fortunately
the residue of this pole vanishes. All is well what ends well.
We can now safely repeat the same argument that leads from eq.(\ref{UU3})
to eq.(\ref{UU5}), and conclude that the various contributions
from the first line of eq.(\ref{UU8}) add up to zero. The second line is
easy to compute, the only contribution comes from the simple pole
at $\zeta=z$:
\beq
= \oint\frac {dz}{2\pi i} \frac 1{z^{n+1}} \oint\frac {d\zeta}{2\pi i}
\oint\frac {dw}{2\pi i} \frac 1{w^{m+1}} \left[
\frac 1{1-zw} - \frac 1{g(w)} \frac {1+w^2}{w}\right]=
\delta_{nm},\quad\quad {n,m\geq 1}\label{UU10}
\eeq
This completes the proof of (\ref{UU}).

\acknowledgments

Two of us (C.M. and D.M.) would like to thank L.F.Alday and M.Cirafici for
useful discussions. This review grew out of various talks given by the
authors in Torino (November 2001), Vietri (March 2002, April 2003 ),
Sao Paulo (July 2002),
Ahrenshoop (August 2002), Anacapri (September 2002) and especially from
lectures delivered by L.B. at Kopaonik (September 2002).
This research was supported by the Italian MIUR
under the program ``Teoria dei Campi, Superstringhe e Gravit\`a''.


\end{document}